\documentclass[12pt]{article}

\usepackage{a4wide}
\usepackage{amsmath}
\usepackage{amssymb}
\usepackage{url}
\usepackage{hyperref}
\usepackage{graphicx}
\usepackage{algorithm}
\usepackage{algorithmic}

\newcommand{\cS}{{\cal S}}
\newcommand{\order}[1]{{\cal O}\left(#1\right)}

\newcommand{\cf}{{\em cf.}\ }
\newcommand{\ie}{{\em i.e.}\ }
\newcommand{\eg}{{\em e.g.}\ }
\newcommand{\GeV}{\,\mathrm{GeV}}
\newcommand{\rsep}{R_{\mathrm{sep}}}
\newcommand{\as}{\alpha_s}
\newcommand{\pttilde}{{\tilde p}_t}

\title{\textsf{A practical Seedless Infrared-Safe Cone jet algorithm}}
\author{Gavin P.~Salam and Gr\'egory Soyez\footnote{On leave from the
    PTF group of the University of Li\`ege.} \footnote{Current address:
    Physics Department, Brookhaven National Laboratory, Upton, NY 11973, USA}\\[1.0em]
  LPTHE, \\
  Universit\'e Pierre et Marie Curie -- Paris 6,\\
  Universit\'e Denis Diderot -- Paris 7,\\
  CNRS UMR 7589, 75252 Paris cedex 05, France.  }

\date{}

\begin{document}

\maketitle
\vspace{-10.3cm}

\begin{flushright}
  arXiv:0704.0292 [hep-ph]\\
  April 2007\\
  %{\tiny $ 
  %  $LastChangedDate: 2007-04-04 10:52:42 -0400 (Wed, 04 Apr 2007) $\!\! $}
\end{flushright}
\vspace{7.5cm}

\begin{abstract}
  Current cone jet algorithms, widely used at hadron colliders, take event
  particles as seeds in an iterative search for stable cones. A
  longstanding infrared (IR) unsafety issue in such algorithms is
  often assumed to be solvable by adding extra `midpoint' seeds, but
  actually is just postponed to one order higher in
  the coupling.  A proper solution is to switch to an exact seedless
  cone algorithm, one that provably identifies all stable cones.  The
  only existing approach takes $N2^N$ time to find jets among $N$
  particles, making it unusable at hadron level. This
  can be reduced to $N^2 \ln N$ time, leading to code (SISCone) whose speed
  is similar to that of public midpoint implementations.
  Monte Carlo tests provide a strong cross-check of an analytical
  proof of the IR safety of the new algorithm, and the absence of any
  `$R_\mathrm{sep}$' issue implies a good practical correspondence
  between parton and hadron levels.
  Relative to a midpoint cone, the use of an IR safe seedless
  algorithm leads to modest changes for inclusive jet spectra, mostly
  through reduced sensitivity to the underlying event, and significant
  changes for some multi-jet observables.

  \vspace*{1.0cm}
  
  \noindent SISCone, the \verb:C++: implementation of the algorithm,
  is available at\\
  \indent\url{http://projects.hepforge.org/siscone/} (standalone),\\
  \indent%
  \url{http://www.lpthe.jussieu.fr/~salam/fastjet/} %
  (FastJet plugin).

\end{abstract}

\newpage
\tableofcontents

%----------------------------------------------------------------------

\section{Introduction}

Two broad classes of jet definition are generally advocated \cite{TeV4LHC}
for hadron colliders.
One option is to use sequential recombination jet algorithms, such as the
$k_t$ \cite{Kt} and Cambridge/Aachen algorithms \cite{Cam}, which
introduce a distance measure between particles, and repeatedly
recombine the closest pair of particles until some stopping criterion
is reached. While experimentally these are starting to be investigated
\cite{D0kt,CDFkt}, the bulk of measurements are currently carried out
with the other class of jet definition, cone jet algorithms (see \eg
\cite{RunII-jet-physics}). In general there are
indications~\cite{SeymourTevlin} that it may be advantageous to use
both sequential recombination and cone jet algorithms because of
complementary sensitivities to different classes of non-perturbative
corrections.

Cone jet algorithms are inspired by the idea \cite{StermanWeinberg} of
defining a jet as an angular cone around some direction of dominant energy
flow. To find these directions of dominant energy flow, cone
algorithms usually take some (or all) of the event particles as
`seeds', \ie trial cone directions. Then for each seed they establish
the list of particles in the trial cone, evaluate the sum of
their 4-momenta, and use the resulting 4-momentum as a new trial
direction for the cone. This procedure is iterated until the cone
direction no longer changes, \ie until one has a ``stable cone''.

Stable cones have the property that the cone axis $a$ (a four-vector)
coincides with the (four-vector) axis defined by the total momentum of
the particles contained in the cone,
\begin{equation}
  \label{eq:stable}
  D\left( p_\mathrm{in\;cone}, a\right) = 0\,,\quad \mathrm{with}\quad
  p_\mathrm{in\;cone} = \sum_{i} p_i\, \Theta(R-D({p}_i,{a}))\,,
%  D\left( \sum_{i} p_i \Theta(R-D({p}_i,{a}))\,,\; a\right) = 0 \,,
\end{equation}
where $D(p,a)$ is some measure of angular distance between the
four-momentum $p$ and the cone axis $a$, and $R$ is the
given opening (half)-angle of the cone, also referred to as the cone
radius.  Typically one defines $D^2(p,a) = (y_p - y_a)^2 + (\phi_p -
\phi_a)^2$, where $y_p, y_a$ and $\phi_p,\phi_a$ are respectively the
rapidity and azimuth of $p$ and $a$.

Two types of problem arise when using seeds as starting points of an
iterative search for stable cones. On one hand, if one only uses
particles above some momentum threshold as seeds, then the procedure
is collinear unsafe.  Alternatively if any particle can act as a seed
then one needs to be sure that the addition of an infinitely soft particle cannot
lead to a new (hard) stable cone being found, otherwise the procedure
is infrared (IR) unsafe.

The second of these problems came to fore in the 1990's
\cite{midpoint}, when it was realised that there can be stable cones
that have two hard particles on opposing edges of the cone and no
particles in the middle, \eg for configurations such as
\begin{equation}
  \label{eq:needsMidpoint}
  p_{t1} > p_{t2};\quad  R < D(p_1,p_2) < (1+p_{t2}/p_{t1})R.
\end{equation}
In traditional iterative cone algorithms, $p_1$ and $p_2$ each act as
seeds and two stable cones are found, one centred on $p_1$, the other
centred on $p_2$. The third stable cone, centred between $p_1$ and
$p_2$ (and containing them both) is not found. If, however, a soft
particle is added between the two hard particles, it too acts as a
seed and the third stable cone is then found. The set of stable cones
(and final jets) is thus different with and without the soft particle
and there is a resulting non-cancellation of divergent real soft
production and corresponding virtual contributions, \ie the algorithm
is infrared unsafe.

Infrared unsafety is a serious issue, not just because it makes it
impossible to carry out meaningful (finite) perturbative calculations, but also
because it breaks the whole relation between the (Born or low-order) partonic
structure of the event and the jets that one observes, and it is
precisely this relation that a jet algorithm is supposed to codify: 
it makes no sense for the structure of multi-hundred GeV jets to
change radically just because hadronisation, the underlying event or
pileup threw a 1 GeV particle in between them.

A workaround for the above IR unsafety problem was proposed in
\cite{midpoint}: after finding the 
stable cones that come from the true seed particles, add 
artificial ``midpoint'' seeds between pairs of stable cones and search
for new stable cones that arise from the midpoint seeds.
For configurations with two hard particles, the midpoint fix
resolved the IR unsafety issue. It was thus adopted as a recommendation
\cite{RunII-jet-physics} for Run~II of the Tevatron and is now coming
into use experimentally~\cite{D0Zjets,CDFConeJetsPaper}.

Recently, it was observed \cite{TeV4LHC} that in certain triangular
three-point configurations there are stable cones that are not
identified even by the midpoint procedure. While these can be
identified by extended midpoint procedures (\eg midpoints between
triplets of particles) \cite{PxCone,CDFcode}, in this article
(section~\ref{sec:linear-IR-unsafety}) we show that there exist yet
other 3-particle configurations for which even this fix does not find
all stable cones.

Given this history of infrared safety problems being fixed and new
ones being found, it seems to us that iterative\footnote{A more
  appropriate name might be the \emph{doubly iterative} cone
  algorithm, since as well as iterating the cones, the cone
  algorithm's definition has itself seen several iterations since its
  original introduction by UA1 in 1983 \cite{UA1}, and even since the
  Snowmass accord~\cite{Snowmass}, the first attempt to formulate a
  standard, infrared and collinear-safe cone-jet definition, over 15
  years ago.} %
cone algorithms should be abandoned. Instead we believe that cone jet
algorithms should solve the mathematical problem of demonstrably finding
all stable cones, \ie all solutions to eq.~(\ref{eq:stable}). This
kind of jet algorithm is referred to as an exact seedless cone jet algorithm
\cite{RunII-jet-physics} and has been advocated before in~\cite{KOS}.
With an exact seedless algorithm, the addition of one or more soft
particles cannot lead to new hard stable cones being found, because
all hard stable cones have already been (provably) found. Therefore
the algorithm is infrared safe at all orders.

Two proposals exist for approximate implementations of the seedless
jet algorithm \cite{RunII-jet-physics,Volobouev}. They both rely on the
event being represented in terms of calorimeter towers, which is
far from ideal when considering parton or hadron-level events.
Ref.~\cite{RunII-jet-physics} also proposed a procedure for an exact
seedless jet algorithm,
% section 3.3.3
intended for fixed-order calculations, and
implemented for example in the MCFM and NLOJet fixed order (NLO)
codes~\cite{MCFM,NLOJet}.%
\footnote{Section 3.4.2 of \cite{RunII-jet-physics} is the source of
  some confusion regarding nomenclature, because after discussing both
  the midpoint and seedless algorithms, it proceeds to show some
  fixed-order results calculated with the seedless algorithm, but
  labelled as midpoint. Though both algorithms are IR safe up to the
  order that was shown, they would not have given identical
  results.\label{foot:342}}
This method takes a time $\order{N 2^N}$ to find jets among $N$
particles. While perfectly adequate for fixed order calculations
($N\le 4$), a recommendation to extend the use of such seedless cone
implementations more generally would have little chance of being
adopted experimentally: the time to find jets in a single (quiet!)
event containing $100$ particles would approach $10^{17}$ years.

Given the crucial importance of infrared safety in allowing one to
compare theoretical predictions and experimental measurements, and the
need for the same algorithm to be used in both, there is a strong
motivation for finding a more efficient way of implementing the seedless
cone algorithm.
Section~\ref{sec:exact-cone} will show how this can be done, first in
the context of a simple one-dimensional example (sec.~\ref{sec:1d}),
then generalising it to two dimensions ($y$, $\phi$,
sec.~\ref{sec:2d}) with an approach that can be made to run in
polynomial ($N^2 \ln N$) time. As in recent work on speeding up the
$k_t$ jet-algorithm \cite{FastJet}, the key insights will be obtained by
considering the geometrical aspects of the
problem. Section~\ref{sec:split--merge} will discuss aspects of the
split--merge procedure.

In section~\ref{sec:tests} we will study a range of physics and practical
properties of the seedless algorithm.
Given that the split--merge stage is complex and so yet another
potential source of infrared unsafety, we will use Monte Carlo
techniques to provide independent evidence for the safety of the
algorithm,  supplementing a proof given in
appendix~\ref{sec:appendix}. We will examine the speed of our coding
of the algorithm and see that it is as fast as publicly available
midpoint codes. We will also study the question of the relation
between the low-order perturbative characteristics of the algorithm,
and its all-order behaviour, notably as concerns the `$R_{sep}$'
issue~\cite{EHT,TeV4LHC}. Finally we highlight physics contexts where
we see similarities and differences between our seedless algorithm and
the midpoint algorithm.
For inclusive quantities, such as the inclusive jet spectrum,
perturbative differences are of the order of a few percent, increasing
to 10\% at hadron level owing to reduced sensitivity to the underlying
event in the seedless algorithm. For exclusive quantities we see
differences of the order of 
$10-50\%$, for example for mass spectra in multi-jet events.

%----------------------------------------------------------------------
\section{Overview of the cone jet-finding algorithm}
\label{sec:cone-def}

\begin{algorithm}
  \caption{A full specification of a modern cone algorithm, 
    governed by four parameters: the cone radius $R$, the overlap
    parameter $f$, the number of passes $N_\mathrm{pass}$ and a
    minimum transverse momentum in the split--merge step, $p_{t,\min}$.
    Throughout, particles are to be combined by summing their
    4-momenta and distances are to be calculated using the
    longitudinally invariant $\Delta y$ and $\Delta \phi$ distance
    measures (where $y$ is the rapidity).  }
  \label{alg:fullcone}
  \renewcommand{\algorithmiccomment}[1]{[#1]}
  \begin{algorithmic}[1]
    \STATE Put the set of current particles equal to the set of all
    particles in the event.
    \REPEAT 
    \STATE Find {\em all}\, stable cones of radius $R$ (see Eq.
    (\ref{eq:stable})) for the current set of particles, \eg using
    algorithm~\ref{alg:fastcone}, section \ref{sec:computational}.
    \label{alg:fullcone:allstable}
    \STATE For each stable cone, create a protojet from the current
    particles contained in the cone, and add it to the list of
    protojets.
    \STATE Remove all particles that are in stable cones from the list
    of current particles.
    \UNTIL{No new stable cones are found, or one has gone around the
      loop $N_\mathrm{pass}$ times.}
    \STATE Run a Tevatron Run-II type split--merge
    procedure~\cite{RunII-jet-physics}, algorithm~\ref{alg:splitmerge}
    (section \ref{sec:split--merge}), on the full list of protojets,
    with overlap parameter $f$ and transverse momentum threshold
    $p_{t,\min}$.
  \end{algorithmic}
\end{algorithm}

Before entering into technical considerations, we outline the
structure of a modern cone jet definition as algorithm~\ref{alg:fullcone},
largely based on the Tevatron Run-II specification
\cite{RunII-jet-physics}. It is governed by four parameters. The cone
radius $R$ and overlap parameter $f$ are standard and appeared in
previous cone algorithms.  The $N_\mathrm{pass}$ variable is new and
embodies the suggestion in~\cite{TeV4LHC} that one should rerun the
stable cone search to eliminate dark towers~\cite{EHT}, \ie particles
that do not appear in any stable cones (and therefore never appear in
jets) during a first pass of the algorithm, even though they can
correspond to significant energy deposits.  A sensible default is
$N_\mathrm{pass}=\infty$ since, as formulated, the procedure will in any
case stop once further passes find no further stable cones. The
$p_{t,\min}$ threshold for the split--merge step is also an addition
relative to the Run~II procedure, inspired by
\cite{PxCone,SeymourTevlin}. It is discussed in section
\ref{sec:split--merge} together with the rest of the split--merge
procedure and may be set to zero to recover the original Run~II
type behaviour, a sensible default.

The main development of this paper is the specification of how to
efficiently carry out step \ref{alg:fullcone:allstable} of
algorithm~\ref{alg:fullcone}. In section~\ref{sec:linear-IR-unsafety}
we will 
show that the midpoint approximation for finding stable cones fails to
find them all, leading to infrared unsafety problems.
Section~\ref{sec:exact-cone} will provide a practical solution. Code
corresponding to this algorithm is available publicly under the name of
`Seedless Infrared Safe Cone' (SISCone).

%----------------------------------------------------------------------
\section{IR unsafety in the midpoint algorithm}
\label{sec:linear-IR-unsafety}

Until now, the exact exhaustive identification of all stable cones was
considered to be too computationally complex to be feasible for
realistic particle multiplicities.
Instead,
the Tevatron experiments streamline the search for stable cones with
the so-called 'midpoint algorithm'~\cite{midpoint}. Given a seed, the
latter calculates the total momentum of the particles contained within
a cone centred on the seed, uses the direction of this momentum as a
new seed and iterates until the resulting cone is stable. The initial
set of seeds is that of all particles whose transverse momentum is
above a seed threshold $s$ (one may take $s=0$ to obtain a
collinear-safe algorithm). Then, one adds a new set of seeds given by
all midpoints between pairs of stable cones separated by less than
$2R$ and repeats the iterations from these midpoint seeds.

\begin{figure}
  \centering
  \includegraphics[width=0.8\textwidth]{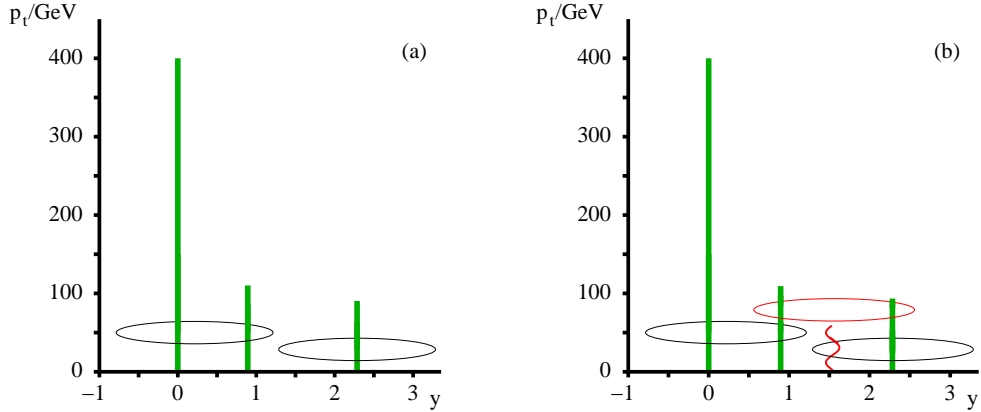}
  \caption{Configuration illustrating one of the IR unsafety problems
    of the midpoint jet algorithm ($R=1$); (a) the stable cones (ellipses)
    found in the midpoint algorithm; (b) with the addition of an
    arbitrarily soft seed particle (red wavy line) an extra stable
    cone is found.}
  \label{fig:example-unsafe}
\end{figure}

The problem with the midpoint cone algorithm can be seen from the
configurations of table~\ref{tab:config}, represented also in
fig.~\ref{fig:example-unsafe}.
\begin{table}
\begin{center}
  \begin{tabular}{c|c|c|c}
    particle & $p_t$ [GeV]& $y$ & $\phi$ \\\hline 
    1        & 400        & 0    &  0 \\
    2        & 110        & 0.9R &  0 \\
    3        &  \;\;90    & 2.3R &  0 \\\hline
    4        &  \;\;\;\;\;\;1.1    & 1.5R &  0 \\
  \end{tabular}
\end{center}
\caption{Particles 1--3 represent a hard configuration. The jets from
  this hard configuration are modified in the midpoint cone algorithm
  when one adds the soft particle $4$.} 
\label{tab:config}
\end{table}
Using particles $1-3$, there exist three stable cones. In a $p_t$-scheme
recombination
procedure (a $p_t$ weighted averaging of $y$ and $\phi$) they are at
$y\simeq\{0.194R, 1.53R, 2.3R\}$.\footnote{In a more standard $E$-scheme
  (four-momentum) recombination procedure the exact numbers depend
  slightly on $R$, but the conclusions are unchanged.}
Note however that starting from particles $1,2,3$ as seeds, one only
iterates to the stable cones at $y\simeq0.194R$ and $y=2.3R$.  Using the
midpoint between these two stable cones, at $y\simeq1.247R$, one iterates
back to the stable cone at $y\simeq 0.194R$, therefore the stable cone at
$y=1.53R$ is never found. The result is that particles $1$ and $2$ are
in one jet, and particle $3$ in another, fig.\ref{fig:example-unsafe}a.

If additionally a soft particle (4) is present to act as a seed near
$y=1.53R$, fig.\ref{fig:example-unsafe}b, then the stable cone there
is found from the iterative procedure. In this case we have three
overlapping stable cones, with hard-particle content $1+2$, $2+3$ and
$3$. What happens next depends on the precise splitting and merging
procedure that is adopted. Using that of \cite{RunII-jet-physics} then
for $f<0.55$ the jets are merged into a single large jet $1+2+3$,
otherwise they are split into $1$ and $2+3$.  Either way the jets are
different from those obtained without the extra soft seed particle,
meaning that the procedure is infrared unsafe. In contrast, a seedless
approach would have found the three stable cones independently of the
presence of the soft particle and so would have given identical sets
of jets.

The infrared divergence arises for configurations with 3 hard
particles in a common neighbourhood plus one soft one (and a further
hard electroweak boson or QCD parton to balance momentum). Quantities
where it will be seen include the NLO contribution to the heavy-jet
mass in $W/Z$+2-jet (or $3$-jet) events, the NNLO contribution to the
$W/Z$+2-jet cross section or the $3$-jet cross section, or
alternatively at NNNLO in the inclusive jet cross section.  The
problem might therefore initially seem remote, since the theoretical
state of the art is far from calculations of any of these quantities.
However one should recall that infrared safety at all orders is a
prerequisite if the perturbation series is to make sense at all. If
one takes the specific example of the $Z$+2-jet cross section
(measured in \cite{D0Zjets}) then the NNLO divergent piece would be
regulated physically by confinement at the non-perturbative scale
$\Lambda_{QCD}$, and would give a contribution of order $\alpha_{EW}
\alpha_s^4 \ln p_t/\Lambda_{QCD}$.  Since $\alpha_s(p_t) \ln
p_t/\Lambda_{QCD} \sim 1$, this divergent NNLO contribution will be of
the same order as the NLO piece $\alpha_{EW} \alpha_s^3$.  Therefore
the NLO calculation has little formal meaning for the midpoint
algorithm, since contributions involving yet higher powers of
$\alpha_s$ will be parametrically as large as the NLO term.%
\footnote{As concerns the measurement~\cite{D0Zjets}, the discussion
  is complicated by the confusion surrounding 
  the nomenclature of the seedless and midpoint algorithms --- while
  it seems that the measurement was carried out with a true midpoint
  algorithm, the calculation probably used the `midpoint' as defined
  in section 3.4.2 of \cite{RunII-jet-physics} (cf.\
  footnote~\ref{foot:342}), which is actually the seedless algorithm,
  \ie the measurements and theoretical predictions are based on
  different algorithms.}  
%Similarly, for jet masses in 3-jet events,
%the infrared divergence at NLO is equivalent to a contribution that is
%parametrically as large as the LO contribution.
The situation for a range of processes is summarised in
table~\ref{tab:failure-cases}.

%  --- indeed the
% magnitude of the effect can be directly estimated from the LO
% difference between results for the mdipoint algorithm and our infrared
% safe one. 

\begin{table}
  \centering
      \begin{tabular}{|l|c|c|}\hline
        Observable        & 1st miss cones at & Last meaningful order \\ \hline
        Inclusive jet cross section &  NNLO & NLO \\ 
        $W/Z/H$ + 1 jet cross section  &  NNLO & NLO \\
        $3$ jet       cross section &   NLO & LO  \\
        $W/Z/H$ + 2 jet cross section &   NLO & LO  \\
        jet masses in $3$~jets, $W/Z/H + 2$~jets   &    LO & none \\\hline
      \end{tabular}
      \caption{Summary of the order ($\as^4$ or $\as^3 \alpha_{EW}$)
        at which stable cones are missed in 
        various processes with a midpoint algorithm, and the corresponding
        last order that can be meaningfully calculated. Infrared
        unsafety first becomes visible one order beyond  that at which
        one misses stable cones.}
  \label{tab:failure-cases}
\end{table}

%======================================================================
\section{An exact seedless cone jet definition}
\label{sec:exact-cone}

One way in which one could imagine trying to `patch' the seed-based iterative
cone jet-algorithm to address the above problem would be to use midpoints
between all pairs of \emph{particles} as seeds, as well as midpoints
between the initial set of stable cones.\footnote{This option was
  actually mentioned in \cite{RunII-jet-physics} but rejected at the
  time as impractical.}  However it seems unlikely that this would
resolve the fundamental problem of being sure that one will
systematically find all solutions of eq.~(\ref{eq:stable}) for any
ensemble of particles.

Instead it is more appropriate to examine exhaustive, non-iterative
approaches to the problem, \ie an exact seedless cone jet algorithm, one
that provably finds all stable cones, as advocated already some time
ago in~\cite{KOS}.

For very low multiplicities $N$, one approach is that
suggested in section~3.3.3 of \cite{RunII-jet-physics} and used in the
MCFM \cite{MCFM} and NLOJet \cite{NLOJet} next-to-leading order codes.
One first identifies all possible subsets of the $N$ particles in the
event.  For each subset $\cS$, one then determines the rapidity
($y_{\cS}$) and azimuth ($\phi_{\cS}$) of the total momentum of the
subset, $p_{\cS} = \sum_{i \in \cS} p_i$ and then checks whether a cone
centred on $y_\cS$, $\phi_\cS$ contains all particles in $\cS$ but no
other particles. If this is the case then $\cS$ corresponds to a
stable cone. This procedure guarantees that all solutions to
eq.~(\ref{eq:stable}) will be found.

In the above procedure there are $\sim 2^N$ distinct subsets of
particles and establishing
whether a given subset corresponds to a stable cone takes time
$\order{N}$.  Therefore the time to identify all stable cones is
$\order{N 2^N}$.  For the values of $N$ ($\le 4$) relevant in
fixed-order calculations, $N2^N$ time is manageable, however as soon
as one wishes to consider parton-shower or hadron-level events, with
dozens or hundreds of particles, $N2^N$ time is prohibitive. A
solution can only be considered realistic if it is polynomial in $N$,
preferably with not too high a power of $N$.

As mentioned in the introduction, approximate procedures for implementing
seedless cone jet algorithms have been proposed in the
past~\cite{RunII-jet-physics,Volobouev}. These rely on considering the
momentum flow into discrete calorimeter towers rather than considering
particles. As such they are not entirely suitable for examining the
full range event levels, which go from fixed-order (few partons), via
parton shower level (many partons) and hadron-level, to detector level
which has both tracking and calorimetry information.

%----------------------------------------------------------------------
\subsection{One-dimensional example}
\label{sec:1d}

To understand how one might construct an efficient exact seedless cone jet
algorithm, it is helpful to first examine a one-dimensional analogue of
the problem. The aim is to identify all solutions to
eq.~(\ref{eq:stable}), but just for (weighted) points on a line. The
equivalent of a cone of radius $R$ is a segment of length $2R$.

Rather than immediately looking for stable segments one instead looks
for all distinct ways in which the segment can enclose a subset of the
points on the line. Then for each separate enclosure one calculates
its centroid $C$ (weighted with the $p_t$ of the particles) and
verifies whether the segment centred on
$C$ encloses the same set of points as the original enclosure. If it
does then $C$ is the centre of a stable segment.

\begin{figure}
  \centering
  \includegraphics[width=0.5\textwidth]{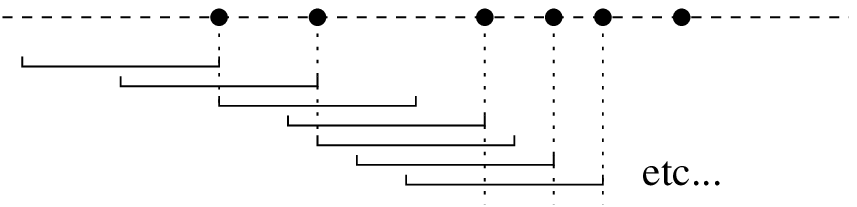}
  \caption{Representation of points on a line and the places where a
    sliding segment has a change in its set of enclosed points.}
  \label{fig:1dSegment}
\end{figure}

A simple way of finding all distinct segment-enclosures is illustrated
in fig.\ref{fig:1dSegment}. First one sorts the points into order on
the line. One then places the segment far to the left and slides it so
that it goes infinitesimally beyond the leftmost point. This is a
first enclosure. Then one slides the segment again until its right
edge encounters a new point or the left edge encounters a contained
point. Each time either edge encounters a point, the point-content of
the segment changes and one has a new distinct enclosure. Establishing
the stability of each enclosure is trivial, since one knows how far
the segment can move in each direction without changing its point
content --- so if the centroid is such that the segment remains  within
these limits, the enclosure corresponds to a stable segment.

The computational complexity of the above procedure, $N \ln N$, is
dominated by the need to sort the points initially: there are
$\order{N}$ distinct enclosures and, given the sorted list, finding the
next point that will enter or leave an edge costs $\order{1}$ time, as
does updating the weighted centroid (assuming rounding errors can be
neglected), so that the time not associated with the sorting step is
$\order{N}$.

%----------------------------------------------------------------------
\subsection{The two-dimensional case}
\label{sec:2d}

%......................................................................
\subsubsection{General approach}

The solution to the full problem can be seen as a 2-dimensional
generalisation of the 
above procedure.\footnote{We illustrate the planar problem rather than
  the cylindrical one since for $R<\pi/2$ the latter is a trivial
  generalisation of the former.} 
The key idea is again that of trying to identify all distinct circular
enclosures, which we also call distinct cones (by `distinct' we mean
having a different point content), and testing the stability of each
one.  In the one-dimensional example there was a single degree of
freedom in specifying the position of the segment and all distinct
segment enclosures could be obtained by considering all segments with
an extremity defined by a point in the set.
In 2 dimensions there are two degrees of freedom in specifying the
position of a circle, and as we shall see, the solution to finding all
distinct circular enclosures will be to examine all circles whose
circumference lies on a \emph{pair} of points from the set.

\begin{figure}
  \centering
  \includegraphics[width=\textwidth]{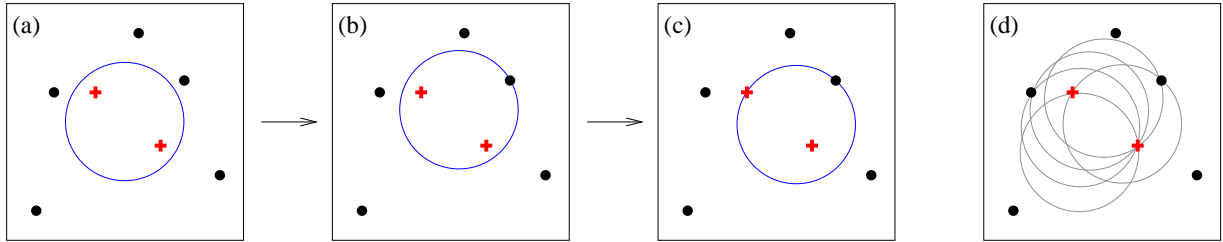}
  \caption{(a) Some initial circular enclosure; (b) moving the circle
    in a random direction until some enclosed or external point
    touches the edge of the circle; (c) pivoting the circle around the
    edge point until a second point touches the edge; (d) all circles
    defined by pairs of edge points leading to the same circular
    enclosure.}
  \label{fig:2dcircle}
\end{figure}

To see in detail how one reaches this conclusion, it is useful to
examine fig.~\ref{fig:2dcircle}. Box (a) shows a circle enclosing two
points, the (red) crosses. Suppose, in analogy with
fig.~\ref{fig:1dSegment} that one wishes to slide the circle until its
point content changes.  One might choose a direction at random and
after moving a certain distance, the circle's edge will hit some point
in the plane, box (b), signalling that the point content is about to
change. In the 1-dimensional case a single point, together with a
binary orientation (taking it to be the left or right-hand point) were
sufficient to characterise the segment enclosure. However in the
2-dimensional case one may orient the circle in an infinite number of
ways. We can therefore pivot the circle around the boundary point.  As
one does this, at some point a second point will then touch the
boundary of the circle, box (c).

The importance of fig.~\ref{fig:2dcircle} is that it illustrates that
for each and every enclosure, one can always move the corresponding
circle (without changing the enclosure contents) into a position where
two points lie on its boundary.%
\footnote{There are two minor exceptions to this: (a) for any point
  separated from all others by more than $2R$, the circle containing
  it can never have more than that one point on its edge --- any such
  point forms a stable cone of its own; (b) there may be
  configurations where three or more 
  points lie on the same circle of radius $R$ (\ie are cocircular) --- given a
  circle defined by a pair of them, the question of which of the
  others is in the circle becomes ambiguous and one should explicitly
  consider all possible combinations of inclusion/exclusion; a
  specific case of this is when there are collinear momenta
  (coincident points), which can however be dealt more simply by
  immediately merging them.\label{ft:exceptions}}
Conversely, if one considers each circle whose boundary is defined by a
pair of points in the set, and considers all four permutations of the edge
points being contained or not in the enclosure, then one will have
identified all distinct circular enclosures.
Note that one given enclosure can be defined by several distinct pairs
of particles, which means that when considering the enclosures defined
by all pairs of particles, we are likely to find each enclosure more
than once, cf.\ fig.~\ref{fig:2dcircle}d.

A specific implementation of the above approach to finding the stable
cones is given as algorithm~\ref{alg:fastcone} below. It runs in expected
time $\order{Nn \ln n}$ where $N$ is the total number of particles and
$n$ is the typical number of particles in a circle of radius $R$.%
\footnote{Given a detector that extends to rapidities $y < y_{\max}$,
  $n/N \sim \pi R^2/(4\pi y_{\max})$, which is considerably smaller
  than $1$ --- this motivates us to distinguish $n$ from $N$.} %
The time is dominated by a step that establishes a traversal order for
the $\order{N n}$ distinct circular enclosures, much as the
one-dimensional ($N \ln N$) example was dominated by the step that
ordered the $\order{N}$ distinct segment enclosures.%
\footnote{For comparison we note that the complexity of public midpoint
algorithm implementations scales as $N^2n$.} %
Some aspects of algorithm~\ref{alg:fastcone} are rather technical and
are explained in the subsubsection that follows. A reader interested
principally in the physics of the algorithm may prefer to skip it on a
first reading.

%----------------------------------------------------------------------
\subsubsection{Specific computational strategies}
\label{sec:computational}

A key input in evaluating the computational complexity of various
algorithms is the knowledge of the number of distinct circular
enclosures (or `distinct cones') and the
number of stable cones. These are both estimated in
appendix~\ref{sec:multiplicities}, and are respectively $\order{N n}$ and
(expected) $\order{N}$.

Before giving the 2-dimensional analogue of the 1-d algorithm of
section~\ref{sec:1d} we examine a simple `brute force' approach
for finding all stable cones. One takes all $\sim Nn$ pairs of points
within $2R$ of each other and for each pair identifies the contents of
the circle and establishes whether it corresponds to a stable cone, at
a cost of $\order{N}$ each time, leading to an overall $N^2 n$ total
cost. This is to be compared to a standard midpoint cone algorithm,
whose most expensive step will be the iteration of the expected
$\order{Nn}$ midpoint seeds, for a total cost also of $N^2 n$,
assuming the average number of iterations from any given seed to be
$\order{1}$.%
\footnote{In both cases one can reduce this to $N n^2$ by tiling the
  plane into squares of edge-length $R$ and restricting the search for
  the circle contents to tiles in the vicinity of the circle centre.}

\begin{algorithm}[tp]
  \caption{Procedure for establishing the list of all stable cones
    (protojets). For simplicity, parts related to the special case of
    multiple cocircular points (see footnote \ref{ft:exceptions}) are
    not shown. They are a straightforward generalisation of steps
    \ref{alg:fastcone:firstcircle} to \ref{alg:fastcone:lastcircle}.}
  \label{alg:fastcone}
  \renewcommand{\algorithmiccomment}[1]{[#1]}
  \begin{algorithmic}[1]
\STATE For any group of collinear particles, merge them into a single particle.
\label{alg:fastcone:collinear}
\FOR{particle $i = 1 \ldots N$ \label{alg:fastcone:for}}
\STATE Find all particles $j$ within a distance $2R$ of $i$. If there
are no such particles, $i$ forms a stable cone of its own. 
\label{alg:fastcone:find-the-j}
\STATE Otherwise
for each $j$ identify the two circles for which $i$ and $j$ lie on the
circumference.  For
each circle, compute the angle of its centre $C$ relative to $i$,
$\zeta = \arctan \frac{\Delta\phi_{iC}}{\Delta y_{iC}}$.
\label{alg:fastcone:find-two-circles}
\STATE Sort the circles found in steps \ref{alg:fastcone:find-the-j}
and \ref{alg:fastcone:find-two-circles} into increasing angle $\zeta$.
\label{alg:fastcone:sort}
\STATE Take the first circle in this order, and call it the current
circle.  Calculate the total momentum and checkxor for the cones that
it defines. Consider all 4 permutations of edge points being included
or excluded. Call these the ``current cones''.
\label{alg:fastcone:firstcircle}
\REPEAT

\FOR{each of the 4 current cones} 
\STATE If this cone has not yet been found, add it to the list of
distinct cones.
\STATE If this cone has not yet been labelled as unstable, establish
if the in/out status of the edge particles (with respect to the cone
momentum axis) is the same as when defining the cone; if it is
not, label the cone as unstable.\label{alg:fastcone:unstable}
\ENDFOR
\STATE Move to the next circle in order. It differs from the previous
one either by a particle entering the circle, or one leaving the
circle. Calculate the momentum for the new circle and corresponding
new current cones by adding (or
removing) the momentum of the particle that has entered (left); the
checkxor can be updated by XORing with the label of that particle.
\label{alg:fastcone:nextcircle}
\UNTIL{all circles considered.
\label{alg:fastcone:lastcircle}
}
\ENDFOR\label{alg:fastcone:candidates}
\FOR{each of the cones not labelled as unstable}
\STATE Explicitly check its stability, and if it is stable, add it to
the list of stable cones (protojets).
\ENDFOR
\end{algorithmic}
\end{algorithm}

One can reduce the computational complexity by using some of
the ideas from the 1-d example, notably the introduction of an
ordering for the boundary points of circles, and the use of the
boundary points as sentinels for instability.
Specifically, three elements will be required:\vspace{-0.5em} 
\begin{itemize}
\item[i)] one needs a way of labelling distinct cones that allows one
  to test whether two cones are the same at a cost of
  $\order{1}$;\vspace{-0.5em}
\item[ii)] one needs a way of ordering one's examination of cones so
  that one can construct the cones incrementally, so as not to pay the
  (at least, see below) $\order{\sqrt{n}}$ construction price anew for
  each cone; \vspace{-0.5em} 
\item[iii)] one needs a way limiting the number of cones for which we
  carry out a full stability test (which also costs at least
  $\sqrt{n}$).
\end{itemize}

To label cones efficiently, we assign a random $q$-bit integer tag to
each particle. Then we define a tag for combinations of particles by
taking the logical exclusive-or of all the tags of the individual
particles (this is easily constructed incrementally and is sometimes
referred to as a checkxor).  Then two cones can be compared by
examining their tags, rather than by comparing their full list of
particles.
With such a procedure, there is a risk of two non-identical cones
ending up with identical tags (`colliding'), which strictly speaking
will make our procedure only `almost exact'.
The probability $p$ of a collision occurring is roughly the square of the
number of enclosures divided by the number of distinct tags. Since we
have $\order{Nn}$ enclosures, this gives $p \sim N^2 n^2/2^q$. By taking $q$
sufficiently large (in a test implementation we have used $q=96$) and using 
a random number generator that guarantees that all bits are decorrelated
\cite{ranlux}, one can ensure a negligible collision probability.%
\footnote{A more refined
  analysis shows that we need only worry about collisions between the
  tags of stable cones and other (stable or unstable) cones --- since
  there are $\order{N}$ stable cones, the actual collision probability
  is more likely to be $\order{Nn^2}/2^q$. In practice for $N\sim
  10^4$ and $n\sim 10^3$ (a very highly populated event) and using
  $q=96$, this gives $ p \sim 10^{-18}$.
  In principle to guarantee an infinitesimal collision probability
  regardless of N, $q$ should scale as $\ln N$, however $N$ will in
  any case be limited by memory use (which scales as $Nn$) so a fixed
  $q$ is not unreasonable.}

Given the ability to efficiently give a distinct label to distinct
cones, one can address points ii) and iii) mentioned above by
following algorithm~\ref{alg:fastcone}. Point (ii) is dealt with
by steps \ref{alg:fastcone:for}--\ref{alg:fastcone:firstcircle}, \ref{alg:fastcone:nextcircle} and \ref{alg:fastcone:lastcircle}: for each particle $i$, one establishes
a traversal order for the circles having $i$ on their edge --- the
traversal order is such that as one works through the circles, the
circle content changes only by one particle at a time, making it easy
to update the momentum and checkxor for the circle.\footnote{Rounding
  errors can affect the accuracy of the momentum calculated this way;
  the impact of this can be minimised by occasionally recomputing the
  momentum of the circle from scratch.} %
One maintains a record of all distinct cones in the form of a hash (as
a hash function one simply takes $\log_2 Nn$ bits of the tag), so
that it only takes $\order{1}$ time to check whether a cone has been
found previously.

Rather than explicitly checking the stability of each distinct cone,
the algorithm examines whether the multiple edge points that define
the cone are appropriately included/excluded in the circle around the
cone's momentum axis, step \ref{alg:fastcone:unstable}. All but a tiny
fraction of unstable 
cones fail this test, so that at the end of
step~\ref{alg:fastcone:candidates} one has a list (of size
$\order{N}$) of candidate stable cones --- at that point one can carry
out a full stability test for each of them.  This therefore deals with
point~(iii) mentioned above.

The dominant part of algorithm~\ref{alg:fastcone} is the ordering of
the circles, step~\ref{alg:fastcone:sort}, which takes $n \ln n$ time
and must be repeated $N$ times. Therefore the overall cost is $N n \ln
n$. As well as computing time, a significant issue is the memory use,
because one must maintain a list of all distinct cones, of which there
are $\order{Nn}$. One notes however that standard implementations of
the split--merge step of the cone algorithm also require $\order{Nn}$
storage, albeit with a smaller coefficient.

It is worth highlighting also an alternative approach, which though
slower, $\order{N n^{3/2}}$, has lower memory consumption and also
avoids the small risk inexactness from the checkxor. It is similar to
the brute-force approach, but uses 2-dimensional computational
geometry tree structures, such as quad-trees \cite{QuadTree} or $k$-d
trees \cite{KDTree}. These involve successive sub-divisions of the
plane (in quadrants, or pairs of rectangles), similarly to what is
done in $1$-dimensional binary trees.
They make it possible to check the stability of a given circle in
$\sqrt{n}$ time (the time is mostly taken by identifying tree cells
near the edge of the circle, of which there are $\order{\sqrt{n}}$),
giving an overall cost of $N n^{3/2}$. The memory use of this form of
approach is $\order{N\smash{\sqrt n}}$, simply the space needed to
store the stable-cone contents.%
\footnote{Though here we are mainly interested in exact approaches,
  one may also examine the question of the speed of the approximate
  seedless approach of Volobouev~\cite{Volobouev}. This approach
  represents the event on a grid and essentially calculates the
  stability of a cone at each point of the grid using a fast-Fourier
  transformation (FFT). In principle, for this procedure to be as good
  as the exact one, the grid should be fine enough to resolve each
  distinct cone, which implies that it should have $\order{N n}$
  points; therefore the FFT will require $\order{N n \ln Nn}$ time,
  which is similar in magnitude to the time that is needed by the
  exact algorithm. An open question remains that of whether
  a coarser grid might nevertheless be `good enough' for many
  practical applications.}

%----------------------------------------------------------------------
\subsection{The split--merge part of the cone algorithm}
\label{sec:split--merge}

\begin{algorithm}[t]
  \caption{The disambiguated, scalar $\pttilde$ based formulation of a
       Tevatron Run-II type split--merge
    procedure~\cite{RunII-jet-physics}, with overlap threshold
    parameter $f$ and transverse momentum threshold $p_{t,\min}$.  To
    ensure boost invariance and IR safety, for the ordering variable
    and the overlap measure, it uses of ${\tilde p}_{t,\mathrm{jet}} =
    \sum_{i\in \mathrm{jet}} |p_{t,i}|$, \ie a scalar sum of the
    particle transverse momenta (as in a `$p_t$' recombination
    scheme).  }
  \label{alg:splitmerge}
  \renewcommand{\algorithmiccomment}[1]{[#1]}
  \begin{algorithmic}[1]
    \REPEAT \label{alg:splitmerge:begin_loop}
    \STATE \label{alg:splitmerge:threshold}
    Remove all protojets with $p_t <p_{t,\min}$.
    \STATE \label{alg:splitmerge:first_jet}
    Identify the protojet ($i$) with the highest ${\tilde p}_{t}$.
    \STATE \label{alg:splitmerge:second_jet} 
    Among the remaining protojets identify the one ($j$) with
    highest ${\tilde p}_{t}$ that shares particles (overlaps) with $i$.
    \IF{there is such an overlapping jet \label{alg:splitmerge:test_overlap}}
    \STATE{Determine the total
      ${\tilde p}_{t,\textrm{shared}} \!=\! \sum_{k\in i
        \& j} |p_{t,k} |$ of the particles shared between $i$ and
      $j$.\label{alg:splitmerge:pt_overlap}}
    \IF{${\tilde p}_{t,\textrm{shared}} < f {\tilde p}_{t,j}$
      \label{alg:splitmerge:start_inner_if}}
    \STATE \label{alg:splitmerge:split} 
    Each particle that is shared between the two protojets is assigned
    to the one to whose axis it is closest. The protojet momenta
    are then recalculated.
    \ELSE%
    \STATE \label{alg:splitmerge:merge} 
    Merge the two protojets into a single new protojet (added
    to the list of protojets, while the two original ones are removed).
    \ENDIF \label{alg:splitmerge:end_inner_if}%
    \STATE If steps
    \ref{alg:splitmerge:start_inner_if}--\ref{alg:splitmerge:end_inner_if}
    produced a protojet that coincides with an existing one, maintain
    the new protojet as distinct from the existing copy(ies).
    \label{alg:splitmerge:disambiguate}
    \ELSE%
    \STATE \label{alg:splitmerge:addjet} 
    Add $i$ to the list of final jets, and remove it from the list of protojets.
    \ENDIF%
    \UNTIL{no protojets are left. \label{alg:splitmerge:end_loop}}
  \end{algorithmic}  
\end{algorithm}
The split--merge part of our cone algorithm is basically that
adopted for Run-II of the Tevatron~\cite{RunII-jet-physics}. It is
shown in detail as algorithm~\ref{alg:splitmerge}. Since it does not
depend on the procedure used to find stable cones, it may largely be
kept as is. We do however include the following small modifications:
\begin{enumerate}
\item The run~II proposal used $E_t$ throughout the split--merge
  procedure. This is not invariant under longitudinal boosts. We
  replace it with $\tilde p_t$, a scalar sum of the transverse momenta
  of the constituents of the protojet. This ensures that the results
  are both boost-invariant and infrared safe. We note that choosing
  instead $p_t$ (a seemingly natural choice, made for example in the
  code of \cite{NLOJet,CDFcode}) would have led to IR unsafety in
  purely hadronic events ---
  the question of the variable to be used for the ordering is actually
  a rather delicate one, and we discuss it in more detail in
  appendix~\ref{sec:split-merge-var}.

\item We introduce a threshold $p_{t,\min}$ below which protojets are
  discarded (step~\ref{alg:splitmerge:threshold} of
  algorithm~\ref{alg:splitmerge}). This parameter is motivated by the
  discussion in \cite{RunII-jet-physics} concerning problems
  associated with an `excess' of stable cones in seedless algorithms,
  notably in events with significant pileup. It provides an infrared
  and collinear safe way of removing the resulting large number of low
  $p_t$ stable cones. By setting it to zero one recovers a behaviour
  identical to that of the Run-II algorithm (modulo the replacement
  $E_t\to \pttilde$, above), and we believe that in practice zero is
  actually a sensible default value.  We note that a similar parameter
  is present in PxCone~\cite{PxCone,SeymourTevlin}.

\item After
  steps~\ref{alg:splitmerge:start_inner_if}--\ref{alg:splitmerge:end_inner_if},
  the same protojet may appear more than once in the list of
  protojets.  For example a protojet may come once from a single
  original stable cone, and a second time from the splitting of
  another original stable cone. The original statement of the
  split--merge procedure~\cite{RunII-jet-physics} did not address this
  issue, and there is a resulting ambiguity in how to proceed.
  One option (as is done for example in the seedless cone code of
  \cite{NLOJet}) is to retain only a single copy of
  any such identical protojets. This however introduces a new source
  of infrared unsafety: an added soft particle might appear in one copy
  of the protojet and not the other and the two protojets would then
  no longer be identical and would not be reduced to a single
  protojet.  This could (and does occasionally, as evidenced in section
  \ref{sec:ir-safety}) alter the subsequent
  split--merge sequence. If one instead maintains multiple identical
  protojets as distinct entities (as is done in the codes of
  \cite{CDFcode,MCFM}), 
  then the addition of a soft particle does not alter the number of
  hard protojet entries in the protojet list and the split--merge part
  of the algorithm remains infrared safe. We therefore choose this
  second option, and make it explicit as
  step~\ref{alg:splitmerge:disambiguate} of
  algorithm~\ref{alg:splitmerge}.
\end{enumerate}
The split--merge procedure is guaranteed to terminate because the
number of overlapping pairs of protojets is reduced
each time an iteration of the loop finds an overlap.
A proof of the infrared safety of this (and the other) parts of our
formulation of the cone algorithm is given in
appendix~\ref{sec:appendix}. 
The computational complexity ($\order{N^2}$) of the split--merge procedure is
generally smaller than that of the stable-cone search, and so we
relegate its discussion to appendix~\ref{sec:split-merge-complexity}.

Finally, before closing this section, let us return briefly to the
top-level of the cone formulation, algorithm~\ref{alg:fullcone} and
the question of the loop over multiple passes. This loop contains just the
stable-cone search, and one might wonder why the split--merge step has
not also been included in the loop.
First consider $p_{t,\min}=0$: protojets found in different passes
cannot overlap, and the split--merge procedure is such that if a
particle is in a protojet then it will always end up in a jet.
Therefore it is immaterial whether the split--merge step is kept
inside or outside the loop. The advantage of keeping it outside the
loop is that one may rerun the algorithm with multiple overlap values
$f$ simply by repeating the split--merge step, without repeating the
search for stable cones. For $p_{t,\min} \neq 0$ the positioning of
the split--merge step with respect to the $N_\mathrm{pass}$ loop would
affect the outcome of the algorithm if all particles not found in
first-pass jets were to be inserted into the second pass stable-cone
search. Our specific formulation constitutes a design choice, which
allows one to rerun with different values of $f$ and $p_{t,\min}$
without repeating the stable-cone search.

%----------------------------------------------------------------------
\section{Tests and comparisons}
\label{sec:tests}

%......................................................................
\subsection{Measures of IR (un)safety}
\label{sec:ir-safety}

In section~\ref{sec:exact-cone} we presented a procedure for finding
stable cones that is explicitly IR safe. In
appendix~\ref{sec:appendix} we provide a proof of the IR safety of the
rest of the algorithm. The latter is rather technical and not short, 
and while we have every reason to believe it to be correct, 
we feel that there is value in supplementing it with complementary 
evidence for the IR safety of the algorithm. As a byproduct, we will 
obtain a measure of the IR unsafety of various commonly used 
formulations of the cone algorithm.

To verify the IR safety of the seedless cone algorithm, we opt for a
numerical Monte Carlo approach, in analogy with that  used in
\cite{caesar} to test the more involved \emph{recursive} infrared
and collinear safety (a prerequisite for certain kinds of resummation).
The test proceeds as follows. One generates a `hard' event consisting
of some number of randomly distributed momenta of the order of some
hard scale $p_{t,H}$, and runs the jet algorithm on the hard event. One
then generates some soft momenta at a scale $p_{t,S} \ll p_{t,H}$,
adds them to the hard event (randomly permuting the order of the
momenta) and reruns the jet algorithm. One verifies that the hard jets
obtained with and without the soft event are identical. If they are
not, the jet algorithm is IR unsafe. For a given hard event one repeats
the test with many different add-on soft events so as to be reasonably
sure of identifying most hard events that are IR unsafe. One then
repeats the whole procedure for many hard events.

\begin{table}
  \centering
  \begin{minipage}{1.0\linewidth}
    \begin{tabular}{l|l|l|c}
      Algorithm & Type                 & IR unsafe & Code \\\hline
      JetClu     & Seeded, no midpoints & 2h+1s \cite{midpoint} & \cite{CDFcode}\\
      SearchCone & Seeded, search cone~\cite{EHT}, midpoints   & 2h+1s
      \cite{TeV4LHC}  & \cite{CDFcode}\\
      MidPoint   & Seeded, midpoints (2-way) & 3h+1s \cite{TeV4LHC} & \cite{CDFcode}\\
      MidPoint-3 & Seeded, midpoints (2-way, 3-way) & 3h+1s & \cite{CDFcode}\\
      PxCone     & Seeded, midpoints ($n$-way),  non-standard SM & 3h+1s & \cite{PxCone}\\
      Seedless [SM-$p_t$]  & Seedless, SM uses $p_t$
      & 4h+1s\footnote{Failures on 4h+1s arise only for $R > \pi/4$;
        for smaller $R$, failures arise only for higher multiplicities} & [here]\\
      Seedless [SM-MIP]  & Seedless, SM merges identical protojets 
      & 4h+1s\footnote{Failures for 4h+1s are extremely rare, but
        become more common for 5h+1s and beyond} & [here]\\
      Seedless [SISCone]  & Seedless, SM of algorithm~\ref{alg:splitmerge}
       & no & [here]\\
    \end{tabular}
  \end{minipage}
  \caption{Summary of the various cone jet algorithms and the code used for
    tests here; \mbox{SM} stands for ``split--merge''; $N$h+$M$s
    indicates that infrared unsafety is revealed with configurations
    consisting of $N$ hard particles and $M$ soft ones, not counting an 
    additional hard, potentially non-QCD, particle to conserve momentum. 
    All codes have been used in the form of plugins to FastJet
    (v2.1)~\cite{FastJet}.}
  \label{tab:cone_summary}
\end{table}

The hard events are produced as follows: we choose a linearly
distributed random number of momenta (between 2 and 10) and for each
one generate a random $p_t$ (linearly distributed,
$2^{-24}p_{t,H}\!\le\! p_t \!\le\! p_{t,H}$, with $p_{t,H}=1000\GeV$), a
random rapidity (linearly distributed in $-1.5 \!<\! y \!<\! 1.5$) and
a random $\phi$. For each hard event we also choose random parameters
for the jet algorithm, so as to cover the jet-algorithm parameter space
($0.3\!<\!R\!<\!1.57$, $0.25\!<\!f\!<\!0.95$, linearly distributed,
the upper limit on $R$ being motivated by the requirement that
$R<\pi/2$; the $p_{t,\min}$ on protojets is set to 0 and the number of
passes is set to $1$). For each
add-on soft event we generate between $1$ and $5$ soft momenta,
distributed as the hard ones, but with the soft scale
$p_{t,S}=10^{-100}\GeV$ replacing $p_{t,H}$. 

We note that the hard events generated as above do not conserve
momentum --- they are analogous to events with a missing energy component
or with identified photons or leptons that are not given as inputs to
the jet clustering. For
the safety studies on the full SISCone algorithm, we therefore also
generate a set of hard events which do have momentum conservation,
analogous to purely hadronic events.

To validate our approach to testing IR safety, we apply it to a range
of cone jet algorithms, listed in table~\ref{tab:cone_summary}, including
the many variants that are IR unsafe.  In PxCone the cut on
protojets is set to $1\GeV$ and in the SearchCone algorithm the search
cone radius is set to $R/2$.

\begin{figure}[tb]
  \centering
  \includegraphics[width=0.5\textwidth,angle=270]{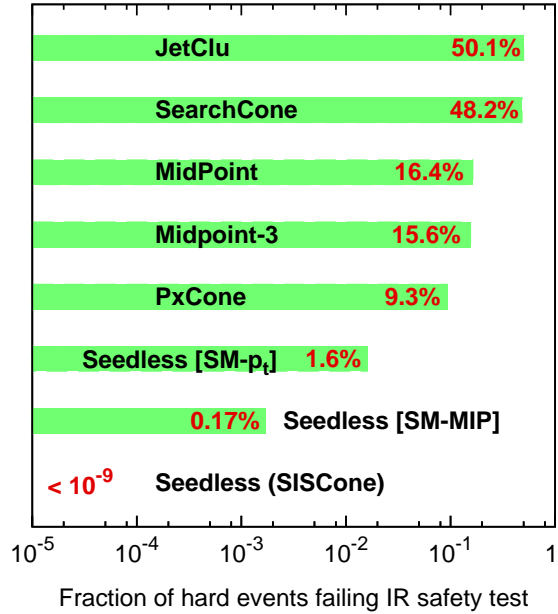}
  \caption{Failure rates for the IR safety tests. The algorithms are
    as detailed in table~\ref{tab:cone_summary}. Seeded algorithms
    have been used with a zero seed threshold. The events used
    do not conserve momentum (\ie have a missing energy
    component), except for the seedless SM-$p_t$ case (where all
    events conserve momentum, to highlight the issue that arises in
    that case) and for SISCone (where we use a mix of momentum
    conserving and non-conserving events so as to fully test the
    algorithm). Further details are given in the text}
  \label{fig:IR_failures}
\end{figure}

The fraction of hard events failing the safety test is shown in
fig.~\ref{fig:IR_failures} for each of the jet algorithms.\footnote{The
  results are based on 80 trial soft add-on events for each hard event
  and should differ by no more than a few percent (relative) from a
  full determination of the IR safety for each hard event (which would
  be obtained in the limit of an infinite number of trial soft add-on
  events for each hard event). For SISCone we
  only use 20 soft add-on events, so as to make it possible to probe a
  larger number of hard configurations.} %
All jet algorithms that are known to be IR unsafe do indeed fail the
tests.
One should be aware that the absolute failure rates depend to some
extent on the way we generated the hard events, and so are to be
interpreted with caution.  Having said that, our hard events have a
complexity similar to the Born-level (lowest-order parton-level) of
events that will be studied at LHC, for
example in the various decay channels of $t\bar t H$ production, and
so both the order of magnitudes of the failure rates and their
relative sizes should be meaningful.

Algorithms that fail on `2h+1s' events have larger failure rates than
those that fail on `3h+1s' events, as would be expected --- they
are `more' infrared unsafe. One notes the significant failure rates
for the midpoint algorithms, $\sim 16\%$, and the fact that adding
3-way midpoints (\ie between triplets of stable cones) has almost no
effect on the failure rate, indicating that triangular configurations
identified as IR unsafe in \cite{TeV4LHC} are much less important than
others such as that discussed in section~\ref{sec:linear-IR-unsafety}.
PxCone's smaller failure rate seems to be due not to its multi-way
midpoints, but rather to its specific split--merge procedure which
leads to fewer final jets (so that one is less sensitive to missing
stable cones).

Seedless algorithms with problematic split--merge procedures lead to
small failure rates (restricting one's attention to small values of
$R$, these values are further reduced). One might be tempted to argue
that such small rates of IR safety failure are unlikely to have a
physical impact and can therefore be ignored. However there is always
a risk of some specific study being unusually sensitive to these
configurations, and in any case our aim here is to provide an
algorithm whose IR safety is exact, not just approximate.

Finally, with a `good' split--merge procedure, that given as
algorithm~\ref{alg:splitmerge}, none of the over $5\times 10^9$ hard
events  tested (a mix both with and without momentum conservation)
failed the IR safety test.
For completeness, we have carried out limited tests also for
$N_\mathrm{pass} = \infty$ and with a $p_{t,\min}$ on protojets of
$100\GeV$, and have additionally performed tests with a larger range
of rapidities ($|y| <3$), collinearly-split momenta, cocircular
configurations, three scales instead of two scales and again found no
failures.
These tests together with the proof given in
appendix~\ref{sec:appendix} give us a good degree of confidence that
the algorithm truly is infrared safe, hence justifying its name.

%......................................................................
\subsection{Speed}
\label{sec:speed}

As can be gathered from the discussion in~\cite{RunII-jet-physics},
reasonable speed is an essential requirement if a new variant of cone
jet algorithm is to be adopted. To determine the speed of various cone
jet algorithms, we use the same set of events taken for testing the
FastJet formulation of the $k_t$ jet algorithm in \cite{FastJet} ---
these consist of a single Pythia \cite{Pythia} dijet event (with
$p_{t,\mathrm{jets}} \simeq 50 \GeV$) to which we add varying numbers
of simulated minimum bias events so as to vary the multiplicity $N$.
Thus the event structure should mimic that of LHC events with pileup.

\begin{figure}[ht]
  \centering
  \includegraphics[width=0.7\textwidth]{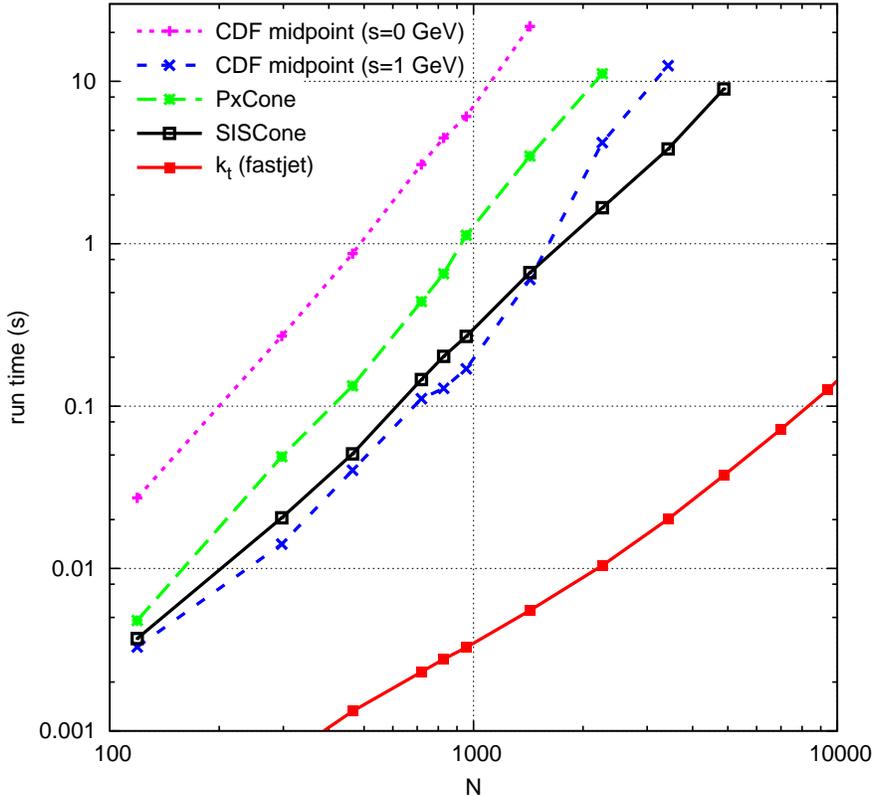}
  \caption{Time to cluster $N$ particles, as a function of $N$, for
    various algorithms, with $R=0.7$ and $f=0.5$, on a 3.4GHz
    Pentium\,{\textregistered} IV processor. For the CDF midpoint
    algorithm, $s$ is the threshold transverse momentum above which
    particles are used as seeds.}
  \label{fig:speed}
\end{figure}

Figure~\ref{fig:speed} shows the time needed to find jets in one event
as a function of $N$. Among the seeded jet algorithms we consider only
codes that include midpoint seeds. For the (CDF) midpoint code
\cite{CDFcode}, written in \texttt{C++}, there is an option of using
only particles above a threshold $s$ as seeds and we consider both the
common (though collinear unsafe) choice $s=1\GeV$ and the (collinear
safe but IR unsafe) $s=0\GeV$. The PxCone code~\cite{PxCone}, written in
Fortran~77, has no seed threshold.

Our seedless code, SISCone, is comparable in speed to the fastest of
the seeded codes, the CDF midpoint code with a seed threshold
$s=1\GeV$, and is considerably faster than the codes without a seed
threshold (not to mention existing exact seedless codes which take
$\sim 1\;$s to find jets among $20$ particles and scale as $N 2^N$).
Its run time also increases more slowly with $N$ than that of the
seeded codes, roughly in agreement with the expectation of SISCone going
as $Nn\ln n$ (with a large coefficient) while the others go as $N^2
n$.  The midpoint code with $s=1\GeV$ has a more complex
$N$-dependence presumably because we have run the timing on a single
set of momenta, and the proportionality between the number of seeds
and $N$ fluctuates and depends on the event structure.

For comparison purposes we have also included the timings for the
FastJet (v2) $k_t$ implementation, which for these values of $N$ uses
a strategy that involves a combination of $N\ln N$ and $Nn$
dependencies.  Timings for the FastJet implementation of the
Aachen/Cambridge algorithm are similar to those for the $k_t$
algorithm.

%......................................................................
\subsection[$\rsep$: an
  inexistent problem]{$\boldsymbol{\rsep}$: an
  inexistent problem}
\label{sec:rsep}

Suppose we have two partons separated by $\Delta R$ and with
transverse momenta $p_{t1}$ and $p_{t2}$ ($p_{t1} > p_{t2}$). Both
partons end up in the same jet if the cone containing both is stable,
\ie if 
\begin{equation}
  \label{eq:rsep-basis}
  \frac{\Delta R}{R} < 1 + z\,,\qquad\quad z = \frac{p_{t2}}{p_{t1}} \,,
\end{equation}
where the result is exact for small $R$ or with $p_t$-scheme
recombination.  Equivalently one can write the probability for two
partons to be clustered into a single jet as
\begin{equation}
  \label{eq:P2to1}
  P_{2\to 1}(\Delta R, z) = \Theta\left(1 + z - \frac{\Delta R}{R}
  \right)\,.
\end{equation}
The limit on $\Delta R/R$ ranges from $1$ for $z = 0$ to $2$ for
$z=1$. This $z$-dependent limit is the main low-order
perturbative difference between the cone algorithm and inclusive
versions of sequential recombination ones like the $k_t$ or
Cambridge/Aachen algorithms, since the latter merge two partons into a
single jet for $\Delta R/R < 1$, independently of their energies.

\begin{figure}[t]
  \centering
  \includegraphics[width=0.6\textwidth]{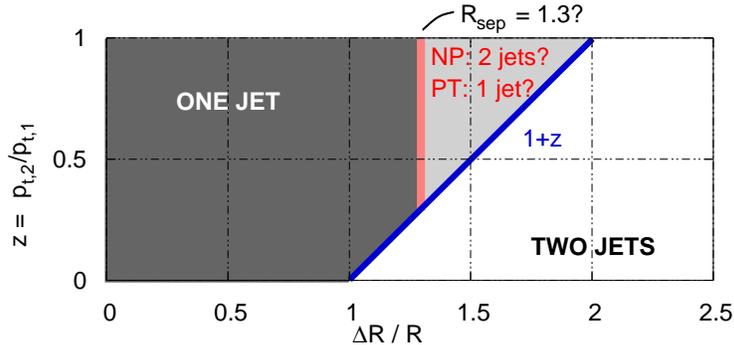}
  \caption{Schematic representation of the phase space region in which
    two partons will end up in a single cone jet versus two jets, at
    the 2-parton level (PT) and, according to the $\rsep$ statement,
    after showering and hadronisation (NP).}
  \label{fig:rsep-schematic}
\end{figure}

A statement regularly made about cone algorithms (see for example
\cite{EHT,TeV4LHC,LHCPrimer}) is that parton showering and
hadronisation reduce the stability of the cone containing the
`original' two partons, leading to a modified `practical' condition
for two partons to end up in a single jet,
\begin{equation}
  \label{eq:rsep-basis-rsep}
  \frac{\Delta R}{R} < \min\left(\rsep\,, 1 + z \right)\,,
\end{equation}
or equivalently,
\begin{equation}
  \label{eq:P2to1Rsep}
  P_{2\to 1}(\Delta R, z) = \Theta\left(1 + z - \frac{\Delta R}{R}  \right)
    \Theta\left(\rsep - \frac{\Delta R}{R}\right) \,,
\end{equation}
with $\rsep \simeq
1.3$~\cite{Abe:1991ui,Abbott:1997fc}.\footnote{The name $\rsep$ was
  originally introduced~\cite{EKSRsep} in the context of NLO
  calculations of hadron-collider jet-spectra, but with a different
  meaning --- there it was intended as a free parameter to model the
  lack of knowledge about the details of the definition of the cone
  jet algorithm used experimentally. This is rather different from the
  current use as a parameter intended to model our inability to
  directly calculate the impact of higher-order and non-perturbative
  dynamics of QCD in cone algorithms.}
This
situation is often represented as in figure~\ref{fig:rsep-schematic},
which depicts the $\Delta R$,~$z$ plane, and shows the
regions in which two partons are merged into one jet or resolved as
two jets. The boundary $\Delta R = 1+z$ corresponds to
eq.~(\ref{eq:rsep-basis}), while the alternative boundary at $\Delta R =
\rsep$ is eq.~(\ref{eq:rsep-basis-rsep}).

So large a difference between the low-order partonic expectation and
hadron-level results would be quite a worrying feature for a jet
algorithm --- after all, the main purpose of a jet algorithm is to
give as close a relation as possible between the first couple of
orders of perturbation theory and hadron level.\footnote{The apparent
  lack of correspondence is considered sufficiently severe that in
  some publications (\eg \cite{CDFConeJetsPaper}) the NLO calculation
  is modified by hand to compensate for this.}

The evidence for the existence of eq.~(\ref{eq:P2to1Rsep}) with
$\rsep=1.3$ seems largely to be
based~\cite{Abe:1991ui,Abbott:1997fc} on merging two events
(satisfying some cut on the jet $p_t$'s), running the jet-algorithm on
the merged event, and examining at what distance particles from the
two events end up in the same jet. This approach indicated that
particles were indeed less likely to end up in the same jet if they
were more than $1.3 R$ apart, however the result is an average over a
range of $z$ values making it hard to see whether 
eq.~(\ref{eq:P2to1Rsep}) is truly representative of the underlying
physics.\footnote{A preliminary version of \cite{LHCPrimer} showed
  more differential results; these, however, seem not to be in the
  definitive version.}

To address the question in more depth we adopt the following
strategy. Rather than combining different events, we use one
event at a time, but with two different jet algorithms.
On one hand we run SISCone with a fairly small value of $R$, $R_{\text{cone}}
= 0.4$. Simultaneously we run inclusive $k_t$ jet-clustering \cite{Kt}
on the
event, using a relatively large $R$ ($R_{k_t} = 1.0$), and identify
any hard $k_t$-jets. For each hard $k_t$ jet we undo its last
clustering step so as to obtain two subjets, $S_1$ and $S_2$ --- these
are taken to be the analogues of the two partons. We then examine whether there is
a cone jet that contains more than half of the $p_t$ of each of $S_1$
and $S_2$. If there is, the conclusion is that the two $k_t$ subjets have
ended up (dominantly) in a single cone jet.

\begin{figure}[tb]
  \centering
  \includegraphics[width=0.7\textwidth]{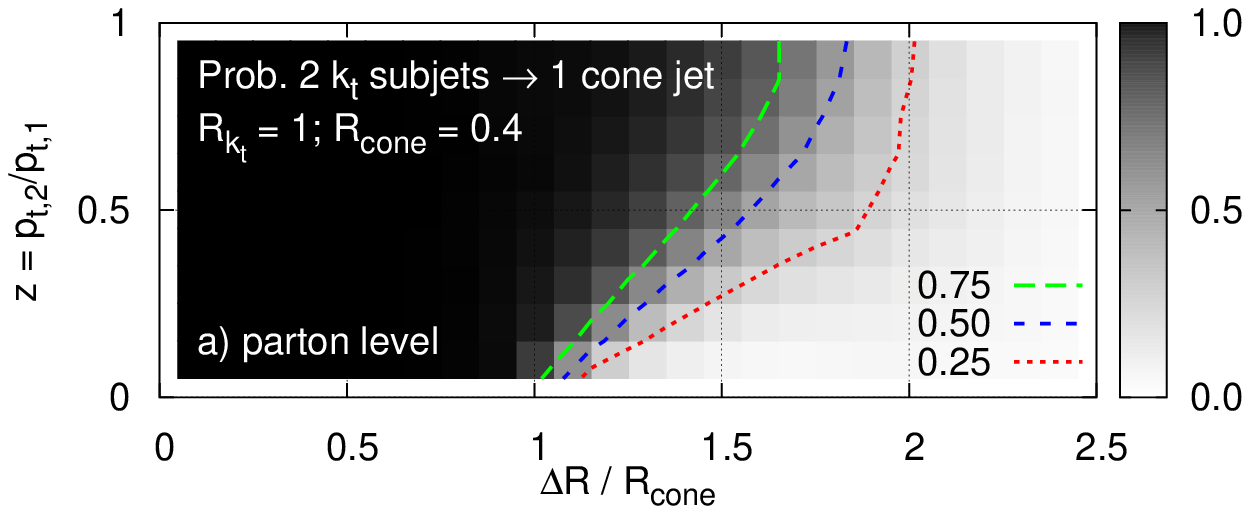}
  \includegraphics[width=0.7\textwidth]{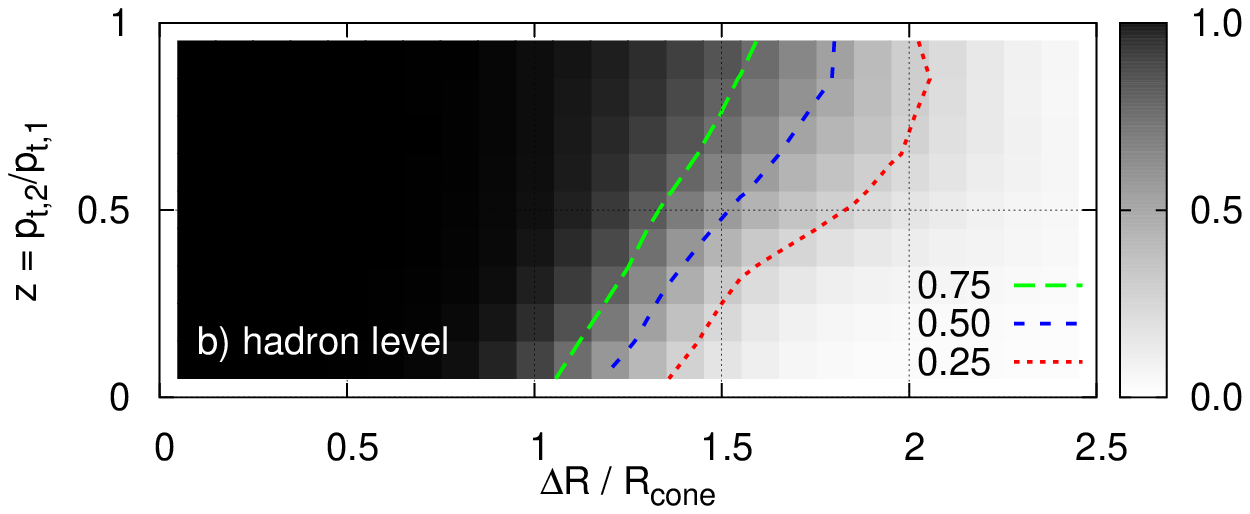}
  \caption{The probability $P_{2\to 1}(\Delta R,z)$ for two
    $k_t$-algorithm subjets to correspond to a single cone jet, as a
    function of $p_{t1}/p_{t2}$ and $\Delta R$ for the two $k_t$
    subjets. Events have been generated with Herwig \cite{Herwig}
    (hadron-level includes the underlying event) and the results are
    based on studying all $k_t$ jets with $p_t > 50\GeV$ and $|y|<$1.
    Further details are to be found in the text.}
  \label{fig:rsep-herwig}
\end{figure}

The procedure is repeated for many events, and one then examines the
probability, $P_{2\to1}(\Delta R, z)$, of the two $k_t$ subjets being
identified with a single cone jet, as a function of the distance
$\Delta R$ between the two subjets, $S_1$ and $S_2$, and the ratio $z$
of their $p_t$'s. The results are shown in fig.~\ref{fig:rsep-herwig}
both at parton-shower level and at hadron level, as simulated with
Herwig~\cite{Herwig}. The middle contour corresponds to a probability
of $1/2$. At parton-shower level this contour coincides remarkably
well with the boundary defined by eq.~(\ref{eq:rsep-basis}), up to
$\Delta R/R=1.7$.  It is definitely not compatible with
eq.~(\ref{eq:rsep-basis-rsep}) with $\rsep=1.3$.
Beyond $\Delta R/R = 1.7$ the contour bends a little and one might
consider interpreting this as an $\rsep \simeq 1.8$.\footnote{Such a
  value has been mentioned to us independently by M.~Wobisch in the
  context of unpublished studies of jet shapes for the SearchCone
  algorithm~\cite{EHT}.} %
However, in that region the transition between $P=1$ and $P=0$ is
broad, and to within the width of the transition, there remains good
agreement with eq.~(\ref{eq:rsep-basis}) --- it seems more natural
therefore to interpret the small deviation from
eq.~(\ref{eq:rsep-basis}) as a Sudakov-shoulder type structure
\cite{Shoulder}, which broadens and shifts the $\Theta$-function of
eq.~(\ref{eq:P2to1}), as would happen with almost any discontinuity in
a leading-order QCD distribution.

Once one includes hadronisation effects in the study,
fig.~\ref{fig:rsep-herwig}b, one finds that the transition region
broadens further, as is to be expected. Now the $P=1/2$ contour shifts
away slightly from 
the $1+z$ result at small $z$ as well.  However, once again this shift
is modest, and of similar size as the breadth of the transition
region.

To verify the robustness of the above results we have examined other
related indicators. One of them is the probability, $P_{2\to2}$ of
finding two cone jets, each containing more than half of the
transverse momentum of just one of the $k_t$ subjets. At two-parton
level, one expects $P_{1\to2} + P_{2\to2} = 1$. Deviation from this
would indicate that our procedure for matching cone jets to $k_t$ jets
is misbehaving. We find that the relation holds to within around $15\%$ over
most of the region, deviating by at most $\sim 25\%$ in a small corner
of phase space $\Delta R / R \simeq 1.5$, $z \simeq 0.2$.
Another test is to examine the fraction $F_{2}$ of the softer $S_2$'s
transverse momentum that is found in the cone that overlaps dominantly
with $S_{1}$. At two-parton level this should be equal to $P_{2\to1}$,
but this would not be the case after showering if there were
underlying problems with our matching procedure. We find however that
$F_{2}$ does agree well with $P_{2\to 1}$. These, together with yet
further tests, lead to us to believe that conclusions drawn from
fig.~\ref{fig:rsep-herwig} are robust.
Finally, while these results have been obtained within a Monte Carlo
simulation, Herwig, a similar study could equally be well carried
experimentally on real events.

So, in contrast to statements that are often made about the cone jet
algorithm, the perturbative picture of when two partons will recombine,
given by eq.~(\ref{eq:P2to1}), seems to be a relatively good indicator
of what happens even after perturbative radiation and hadronisation.
In particular the evidence that we have presented strongly disfavours
the $\rsep$-based modification, eq.~(\ref{eq:P2to1Rsep}).
This is a welcome finding, and should help provide a firmer basis for
cone-based phenomenology.

%......................................................................
\subsection{Physics impact of seedless v.\ midpoint cone}
\label{sec:seedless_v_midpoint}

%In this section, we discuss the physical impact of the IR-safe SISCone
%algorithm. 

In this section, we discuss the impact on physical measurement of
switching from a midpoint type algorithm to a seedless IR-safe one
such as SISCone.
We study two physical observables,
the inclusive jet spectrum and the jet mass spectrum in 3-jet events.
%We have compared the results given by the seedless algorithm against
%the ones obtained with the midpoint algorithm. 
The spectra have been obtained by generating events with a Monte-Carlo
either at fixed order in perturbation theory (NLOJet \cite{NLOJet}) or
with parton showering and hadronisation (Pythia \cite{Pythia}), and by
performing the jet analysis on each event using three different
algorithms (each with $R=0.7$ and $f=0.5$, and additionally in the
case of SISCone, $N_\mathrm{pass} = 1$ and $p_{t,\min}=0$):
\begin{enumerate}
\item SISCone: the seedless, IR-safe definition described in algorithms
  \ref{alg:fullcone}--\ref{alg:splitmerge}; 
\item midpoint(0): the midpoint algorithm using all particles as seeds;
\item midpoint(1): the midpoint algorithm using as seeds all particles above a threshold of 1 GeV.
\end{enumerate}
We have used a version of the CDF implementation of the midpoint
algorithm modified to have the split--merge step based on $\pttilde$
rather than $p_t$ (so that it corresponds to
algorithm~\ref{sec:split--merge} with $p_{t,\min}=0$).
The motivation for this is that we are mainly interested in the physics
impact of having midpoint versus all stable cones, and the comparison
is simplest if the subsequent split--merge procedure is identical in
both cases.\footnote{We could also have compared SISCone with a
  midpoint algorithm using $p_t$ in the split--merge (a common
  default); the figures we show below would have stayed unchanged at
  the $1\%$ level for the inclusive spectrum, while for the jet masses
  the effects range between a few percent at moderate masses and
  $10-20\%$ in the high-mass tail.}

We shall first present the results obtained for the inclusive jet
spectrum and then discuss the jet mass spectrum in 3-jet events.
Most studies carried out in this section have used kinematics
corresponding to the Tevatron Run II, \ie a centre-of-mass energy
$\sqrt{s}=1.96$ TeV, and usually, for simplicity we have chosen not to
impose any cuts in rapidity.

\subsubsection{Inclusive jet spectrum}
\label{sec:inclspect}
\begin{figure}
\centerline{\includegraphics{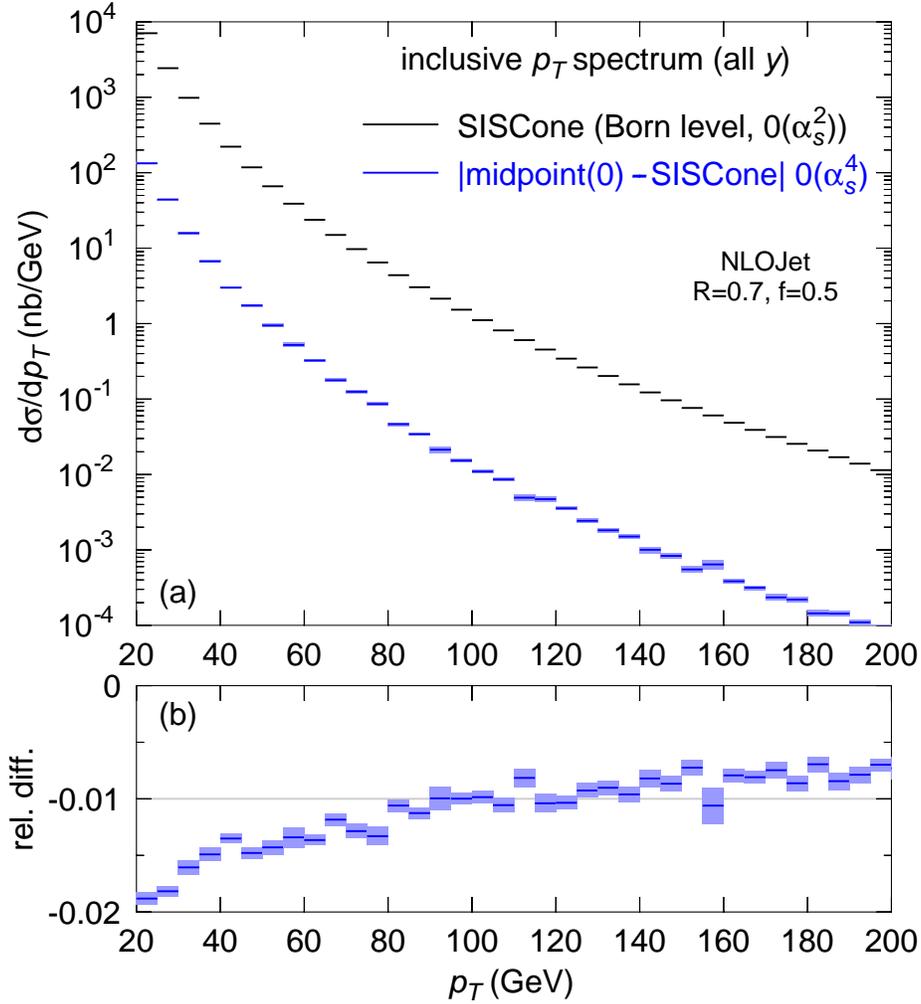}}
\caption{(a) Inclusive jet spectrum: the upper curve gives the
  leading-order ($\order{\alpha_s^2}$) spectrum, while the lower
  (blue) curve gives the difference between the SISCone and
  midpoint(0) algorithm, obtained from the $\order{\alpha_s^4}$
  tree-level amplitude; (b) the relative difference.}
\label{fig:nlojet_incl}
\end{figure}

As discussed in section~\ref{sec:linear-IR-unsafety}, the differences
between the midpoint algorithm and SISCone are expected to start when
we have 3 particles in a common neighbourhood plus one to balance
momentum. For pure QCD processes this corresponds to $2\to 4$
diagrams, $\order{\as^4}$. This is NNLO for the inclusive spectrum.
Though a NNLO calculation of the inclusive spectrum is beyond today's
technology (for recent progress, see \cite{Daleo}), we can easily
calculate the $\order{\as^4}$ difference between midpoint and SISCone,
using just tree-level $2\to4$ diagrams, since the difference between
the algorithms is zero at orders $\as^2$ and $\as^3$, \ie we can
neglect two-loop $2\to2$ diagrams and one-loop $2\to3$ diagrams. The
significance of the difference can be understood by comparing to the
leading order spectrum, which is identical for the two algorithms.

Figure \ref{fig:nlojet_incl} shows the resulting spectra: the upper
plot gives the leading order inclusive spectrum together with the
difference between SISCone and midpoint(0) at $\order{\as^4}$. The
lower plot shows the relative difference. One sees that the use of the
IR-safe seedless cone algorithm introduces modest corrections, of
order 1-2\%, in the inclusive jet spectrum. This order of magnitude is
roughly what one would expect, since the differences only appear at
relative order $\as^2$. As we will see below, larger differences will
appear when one examines more exclusive quantities.
\begin{figure}[tb]
  \centering
  \includegraphics[width=0.48\textwidth]{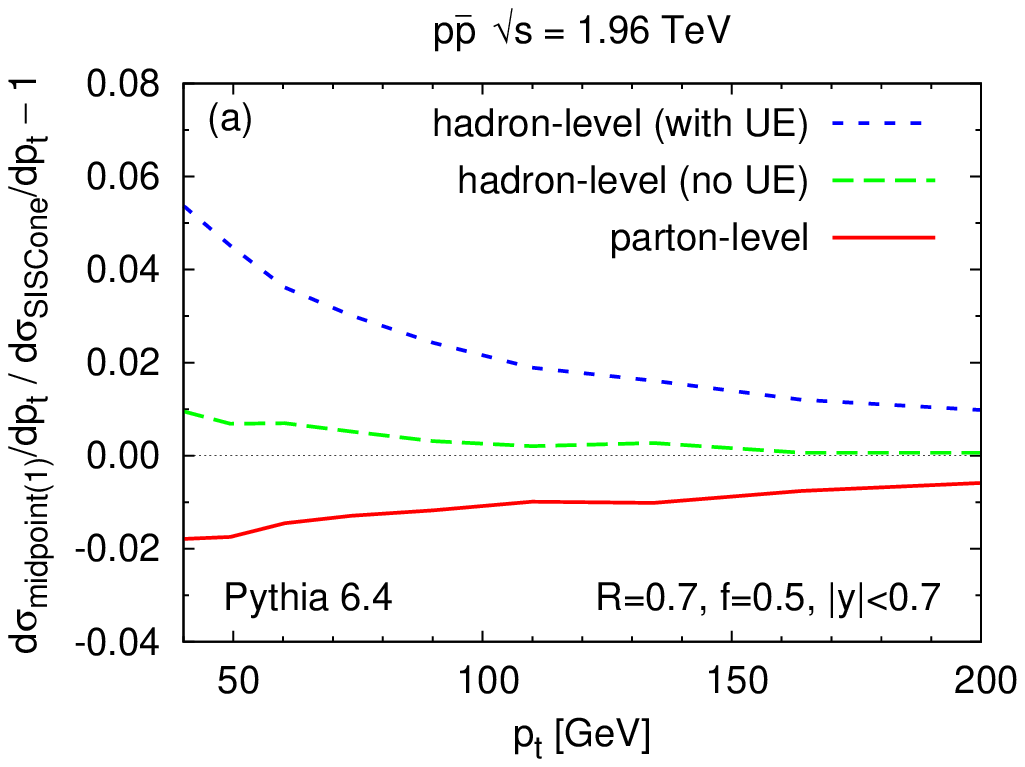}\hfill
  \includegraphics[width=0.48\textwidth]{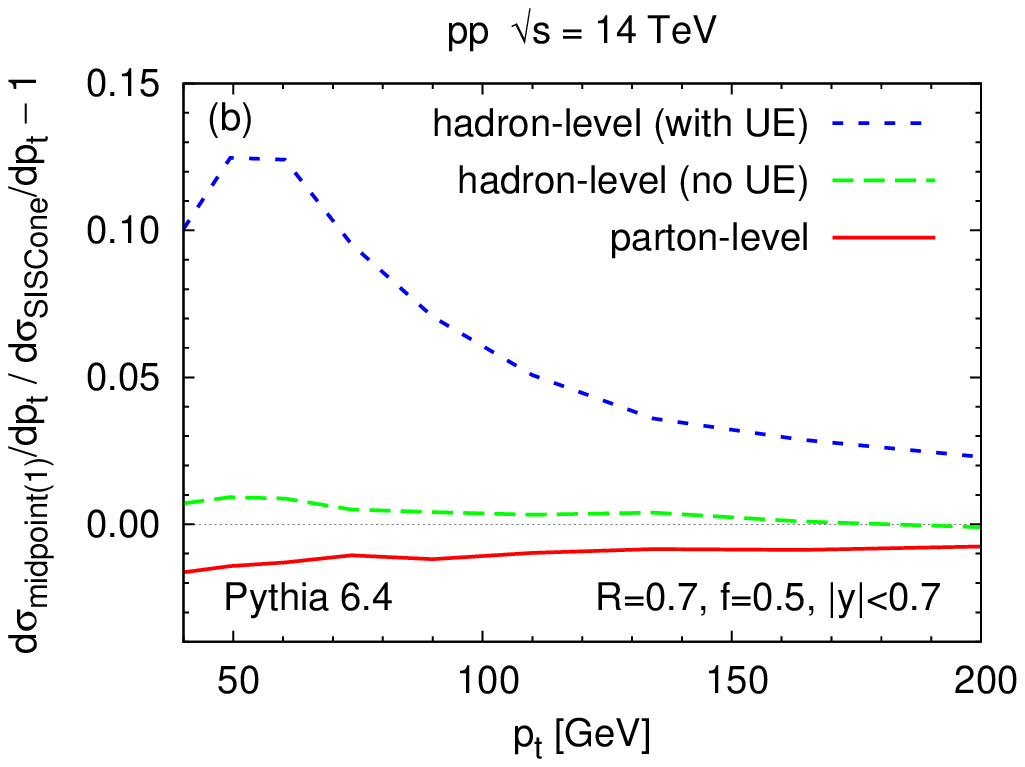}
  \caption{Relative difference between the inclusive jet spectra for
    midpoint(1) and SISCone, obtained from Pythia at parton level,
    hadron level without underlying event (UE) contributions, and
    hadron level with UE. Shown (a) for Tevatron collisions and (b)
    for LHC collisions.}
  \label{fig:incl-3lev}
\end{figure}

In addition, we have used Herwig and Pythia to investigate the
differences between midpoint(1) and SISCone with parton showering.
Both generators give similar results, and we show the results just of
Pythia, fig.~\ref{fig:incl-3lev}a. The difference at parton level is
very similar to what was observed at fixed order. At hadron level
without underlying event (UE) corrections, the difference remains at
the level of $1-2\%$ (though it changes sign); once one includes the
underlying event contributions, the difference increases noticeably at
lower $p_t$ --- this is because the midpoint(1) algorithm receives
somewhat larger UE corrections than SISCone. Since the underlying
event is one of the things that is likely to change from Tevatron to
LHC, in figure \ref{fig:incl-3lev}b we show similar curves for LHC
kinematics.  At parton level and at hadron level without the
underlying event, the results are essentially the same as for the
Tevatron.  With the underlying event included, the impact of the
missing stable cones in the midpoint algorithm reaches of the order of
10 to 15\%, and thus starts to become quite a significant effect. With
Herwig, we find that the impact is little smaller because its
underlying event is smaller than Pythia's at the LHC.

%......................................................................
\subsubsection{Jet masses in 3-jet events}
\label{sec:incl-mass}

\begin{figure}
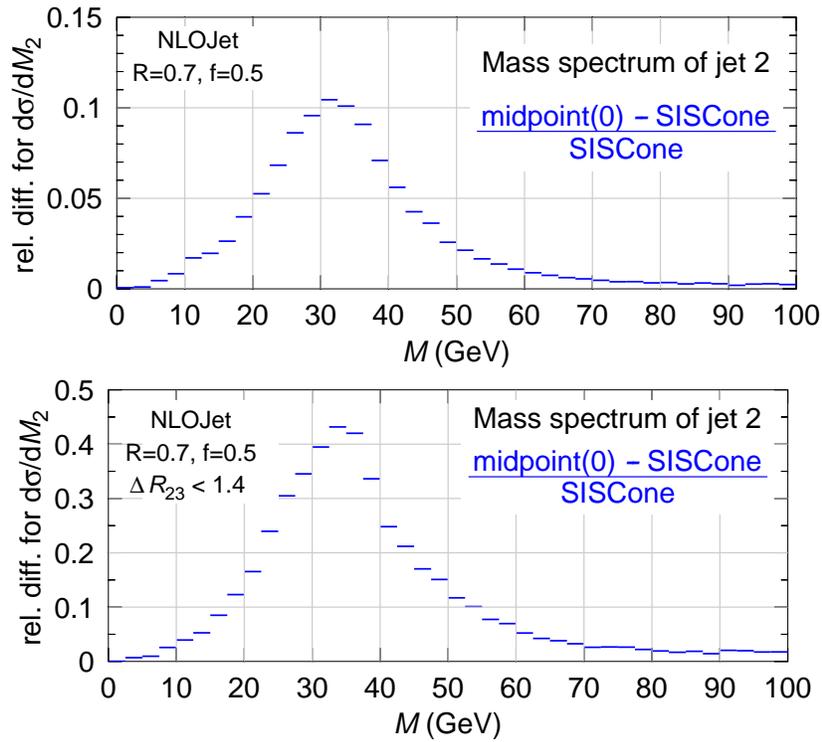

\centerline{\includegraphics[scale=0.9]{nlojet_mass2.ps}}
\centerline{\includegraphics[scale=0.9]{nlojet_mass2_less2R.ps}}
\caption{Mass spectrum of the second hardest jet as obtained with the
  different cone algorithms on tree-level 4-particle events (generated
  with NLOJet): the plots shows the relative difference between the
  midpoint and SISCone results. In the upper plot we consider all
  three-jet events satisfying the transverse-momentum cuts, while in
  the lower plot (note scale) we consider only those in which second
  and third jet are separated by $\Delta R_{23} < 2R$.}
\label{fig:nlojet_mass2}
\end{figure}

\begin{figure}
\centerline{\includegraphics[angle=270,scale=0.8]{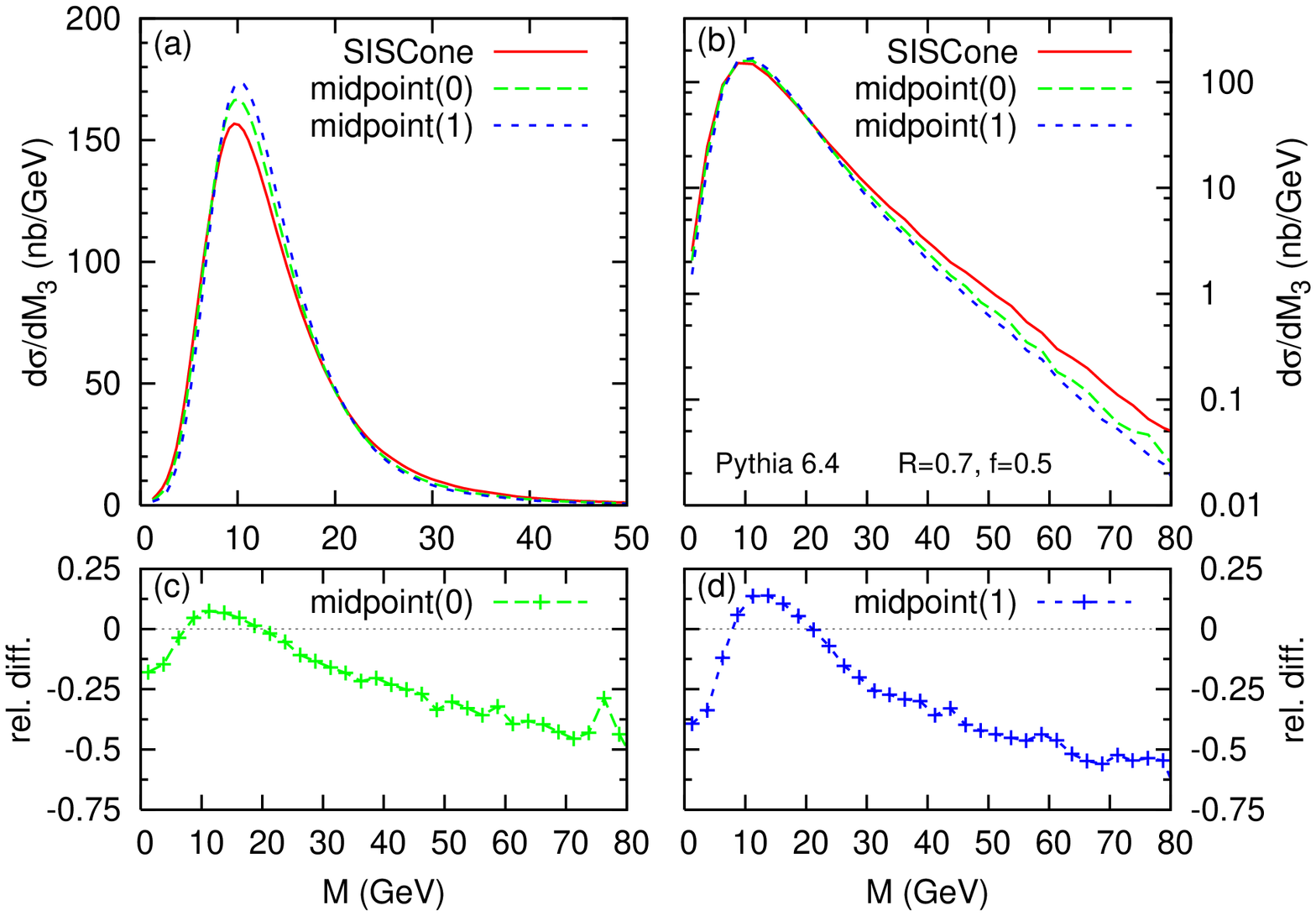}}
\caption{Mass spectrum of the third hardest jet obtained from the
  different cone algorithms run on three-jet Pythia events. The
  top-left (top-right) plot shows the spectrum in linear (logarithmic)
  scale and the bottom plots show the relative difference between each
  midpoint algorithm and SISCone. See the text for the details of the
  event selection.}
\label{fig:pythia_mass}
\end{figure}

As well as the inclusive jet $p_T$ spectrum, we can also study more
exclusive quantities. One example is the jet-mass spectrum in multi-jet
events. Jet-masses are potentially of interest for QCD studies,
particle mass measurements \cite{HoangMantry} and new physics
searches, where they could be used to identify highly boosted W/Z/H
bosons or top quarks produced in the decays of new heavy
particles~\cite{JetMass}.

The simplest multi-jet events in which to study jet masses are 3-jet events.
There, the masses of all the jets vanish at the 3-particle
level. The first order at which the jet masses become non-zero is
$\order{\alpha_s^4}$ and this is also the order at which differences
appear between the midpoint and seedless cone algorithms.
Therefore, as in section~\ref{sec:inclspect}, we generate $2\to4$
tree-level events, but now keep only those
with exactly 3 jets with $p_T\ge 20$ GeV in the final state. We 
further impose that the hardest jet should have a $p_T$ of  at least
120 GeV and the second hardest jet a $p_T$ of at least 60 GeV. With
these cuts we can compute the jet-mass spectrum for each of the
three jets and for the three different algorithms. 

In the upper plot of Figure \ref{fig:nlojet_mass2}, we show the
relative difference ``(midpoint(0) - SISCone)/SISCone'' for the mass
spectrum of the second hardest jet.
In the lower plot we show the same quantity for events in which we
have placed an additional requirement that the $y-\phi$ distance
between the second and third jets be less than $2R$ (such distance cuts
are often used when trying to reconstruct chains of particle decays).
The midpoint algorithm's omission of certain stable cones leads to an
overestimate of the mass spectrum by up to $\sim 10\%$ without a
distance cut (much smaller differences are observed for the first and
third jet) and of over $40\%$ with a distance cut.  The problem is
enhanced by the presence of the distance cut because many more of the
selected events then have three particles in a common neighbourhood,
and this is precisely the situation in which the midpoint algorithm
misses stable cones (\cf section~\ref{sec:linear-IR-unsafety}).

We emphasise also that the NLO calculation of these mass spectra would
be impossible with a midpoint algorithm, because the $10-40\%$
tree-level differences would be converted into an infrared divergent
NLO contribution.

A general comment is that the problems seen here for the midpoint
algorithm without a distance cut are of the same general order of
magnitude as the 16\% failure rate in the IR safety tests of
section~\ref{sec:ir-safety}, suggesting that the absolute failure
rates given there are a good indicator of the degree of seriousness of
issues that can arise in generic studies with the infrared unsafe
algorithms.

In addition to this fixed-order parton-level analysis, we have studied
the jet masses in 3-jet events at hadron level (\ie after parton
showering and hadronisation) using events generated with Pythia. At
hadron level many more seeds are present, due to the large particle
multiplicity. One might therefore expect the midpoint algorithm to
become a good approximation to the seedless one.

For the mass of the second hardest jet, \ie the quantity we studied at
fixed order in
figure~\ref{fig:nlojet_mass2}, we find that the midpoint and seedless
algorithms do give rather similar results at hadron level.  In other
words differences that we see in a leading order calculation are not
propagated through to the full hadron level result.  This is a serious
practical issue for the midpoint algorithm, because a jet algorithm's
principal role is to provide a good mapping between low-order parton
level and hadron level.

Nevertheless, despite the many seeds that are present at hadron level,
we find that 
there are still some observables for which the midpoint algorithm's
lack of stable cones does have a large impact even at hadron level.
This is the case that the mass distribution of the third hardest jet,
shown in figure~\ref{fig:pythia_mass} (obtained without a distance
cut) on both linear and logarithmic scales so as to help visualise
the various regions of the distribution. Moderate differences are
present in the peak region, but in the tail of the distribution they
become large, up to 50\%. They are greater for midpoint(1) than
for midpoint(0), because the seed threshold causes fewer stable cones
to be found with the midpoint(1) algorithm.

These results have been checked using the Herwig Monte-Carlo.
We have observed similar differences at  parton-shower level, at the
hadron level and at the hadron level including underlying event, both
in the peak of the distribution and in the tail. We note
that hadronisation corrections are substantial in the tail of the
distribution, both for the midpoint and SISCone algorithms.

The above results confirm what one might naturally have expected:
while very inclusive quantities may not be overly sensitive to the
deficiencies of one's jet algorithm, as one extends one's
investigations to more exclusive quantities, those deficiencies begin
to have a much larger impact.

%======================================================================
\section{Conclusions}
\label{sec:conclusions}

Given the widespread use of cone jet algorithms at the Tevatron and
their foreseen continued use at LHC, it is crucial that they be defined in
an infrared safe way. This is necessary in general so as to ensure
that low-order parton-level considerations about cone jet-finding hold
also for the fully showered, hadronised jets that are observed in
practice. It is also a prerequisite if measurements are to be
meaningfully compared to fixed order (LO, NLO, NNLO) predictions.

The midpoint iterative cone algorithm currently in use is infrared
unsafe, as can be seen by examining the sets of stable cones that are
found for simple three-parton configurations.  This may seem
surprising given that the midpoint algorithm was specifically designed
to avoid an earlier infrared safety problem --- however the midpoint
infrared problem appears at one order higher in the coupling, and this
is presumably why it was not identified in the original analyses. The
tests shown in section~\ref{sec:ir-safety} suggest that the
midpoint-cone infrared safety problems, while smaller than without the
midpoint, are actually quite significant ($\sim 15\%$).

We therefore advocate that where a cone jet algorithm is used, it be a
seedless variant.
For such a proposal to be realistic it is crucial that the seedless
variant be practical. The approaches adopted in fixed order codes take
$\order{N2^N}$ time and are clearly not suitable in general. Here we
have shown that it is possible to carry out exact seedless jet-finding
in expected $\order{N n^{3/2}}$ time with $\order{N n^{1/2}}$ storage,
or almost exactly\footnote{with a failure probability that can be made
  arbitrarily small and that we choose to be $\lesssim 10^{-18}$.} in
expected $\order{N n \ln n}$ time with $\order{N n}$ storage (we
recall that $N$ is the total number of particles, $n$ the typical
number of particles in a jet). The second of these approaches has been
implemented in a \texttt{C++} code named SISCone, available also as a
plugin for the FastJet package. For $N \sim 1000$ it is comparable in
speed to the existing CDF midpoint code with $1\GeV$ seeds. While this
is considerably slower than the $N \ln N$ and related
FastJet strategies~\cite{FastJet} for the $k_t$ and Cambridge/Aachen jet
algorithms, it remains within the limits of usability and provides for
the first time a cone algorithm that is demonstrably infrared and
collinear safe at all orders, and suitable for use at parton level,
hadron level and detector level.

As well as being infrared safe, a jet algorithm must provide a
faithful mapping between expectations based on low-order perturbative
considerations, and observations at hadron level. There has been
considerable discussion of worrisome possible violations of such a
correspondence for cone algorithms, the ``$\rsep$'' issue. For
SISCone we find however that the correspondence holds well.

An obvious final question is that of the impact on physics results of
switching from the midpoint to the seedless cone. For inclusive
quantities, one expects the seedless cone jet algorithm to give results
quite similar to those of the midpoint cone, because the IR unsafety
of the midpoint algorithm only appears at relatively higher orders.
This is borne out in our fixed order and parton-shower studies of the
inclusive jet spectrum where we see differences between the midpoint
and SISCone algorithms of about a couple of percent. At moderate $p_t$
at hadron level, the differences can increase to $5-10\%$,
because SISCone has a lower sensitivity to the underlying event, a
welcome `fringe-benefit' of the seedless algorithm.

For less inclusive quantities, for example the distribution of jet
masses in multi-jet events, differences can be significant. We find
that for 3-jet events, the absence of some stable cones (\ie infrared
unsafety) in the midpoint algorithm leads to differences compared to
SISCone at the $\sim 10\%$ level at leading order ($\alpha_s^4$) in a
large part of the jet-mass spectrum. Greater effects still, up to
$50\%$, are seen with specific cuts at fixed order, and in
the tails of the jet-mass spectra for parton-shower events.
Thus, even if the infrared safety issues of the midpoint algorithm
appear to be at the limit of today's accuracy when examining inclusive
quantities, for measurements of even moderate precision in multi-jet
configurations (of increasing interest at Tevatron and omnipresent at
LHC), the use of a properly defined cone algorithm such as SISCone is
likely to be of prime importance.

%----------------------------------------------------------------------
\subsection*{Acknowledgements}

We are grateful to Markus Wobisch for many instructive discussions
about cone algorithms, Steve Ellis and Joey Huston for exchanges about
their IR safety and $\rsep$, Matteo Cacciari for helpful suggestions
on the SISCone code and Giulia Zanderighi for highlighting the
question of collinear safety. We thank them all, as well as George
Sterman, for useful comments 
and suggestions on the manuscript. We also gratefully acknowledge
Mathieu Rubin for a careful reading of an early version of the
manuscript, Andrea Banfi for pointing out a relevant reference and
Torbj\"orn Sj\"ostrand for assistance with Pythia.
The infrared unsafe configuration shown here was discovered subsequent
to discussions with Mrinal Dasgupta on non-perturbative properties of
cone jet algorithms.
This work has been supported in part by grant ANR-05-JCJC-0046-01 from
the French Agence Nationale de la Recherche. G.S. is funded by the 
National Funds for Scientific Research (Belgium).
Finally, we thank the Galileo Galilei Institute for Theoretical
Physics for hospitality and the INFN for partial support during the
completion of this work.

%----------------------------------------------------------------------
\begin{appendix}

%----------------------------------------------------------------------
\section{Further computational details}
%----------------------------------------------------------------------
\subsection{Cone multiplicities}
\label{sec:multiplicities}

In evaluating the computational complexity of (computational) algorithms for various
stages of the cone jet algorithm it is necessary to know the numbers of
distinct cones and of stable cones. Such information also constitutes
basic knowledge about cone jet definitions, which may for example be of
relevance in understanding their sensitivity to pileup, \ie multiple
$pp$ interactions in the same bunch crossing.

Since large multiplicities will be due to pileup, let us consider a
simple model for the event structure which mimics pileup, namely a set
of momenta distributed randomly in $y$ and $\phi$ and all with similar
$p_t$'s (or alternatively with random $p_t$'s in some limited range).

Given that the particles will be spread out over a region in $y$, $\phi$
that is considerably larger than the cone area, in addition to $N$,
the total number of particles, it is useful to introduce also $n$, the
number of points likely to be contained in a region of area $\pi R^2$.

The first question to investigate is that of the number of distinct
cones. The number of pairs of points that has to be investigated is
$\order{N n}$. However some of these pairs of points will lead to
identical cones. It is natural to ask whether, despite this, the
number of distinct cones is still $\order{N n}$. To answer this
question, one may examine how far one can displace a cone in any given
direction before its point content changes.  The area swept when
moving a cone a distance $\delta\! R$ is $4R \,\delta R $, and the
average number of points intersected is $4\rho R \,\delta R $ where
$\rho = \order{n/R^2}$ is the density of points (per unit area).
Therefore the distance moved before the cone edge is likely to touch a
point is $\delta R = (4\rho R)^{-1} = \order{R/n}$. Correspondingly
the area in which one can move the centre of cone without changing the
cone's contents is $\pi (\delta R)^2 = \order{R^2/n^2}$. Given that
the total area is $\order{R^2 N/n}$ we have that the number of
distinct cones is $\order{N n}$, the same magnitude as the number of
relevant point pairs.

Let us now consider the number of stable cones. If we take a cone at
random and sum its momenta then the resulting momentum axis will
differ from the original cone axis by an amount typically of order
$R/\sqrt{n}$ (since the standard deviation of $y$ and $\phi$ for set
of points in the cone is $\order{R}$). The probability of the
difference being $\lesssim R/n$ in both the $y$ and $\phi$ directions
(\ie the probability that the new axis contains the same set of
particles) is $\sim (R/n)^2 / (R/\sqrt{n})^2 \sim 1/n$. Therefore the
number of stable cones is $\order{N}$.
This assumes a random distribution of particles. There may exist
special classes of configurations for which the number of stable cones
is greater than $\order{N}$. Therefore timing results that are
sensitive to the number of stable cones are to be understood as
``expected'' results rather than rigorous upper bounds.

%----------------------------------------------------------------------
\subsection{Computational complexity of the split--merge step}
\label{sec:split-merge-complexity}

To study the computational complexity of the split--merge step,
we work with the expectation that there are $\order{N}$ initial
protojets (as discussed above) and that there will be roughly $N/n \ll
N$ final jets (since there are $\order{n}$ particles per jet).  It is
reasonable to assume that there will be roughly equal numbers of merging and
splitting operations. Splitting leaves the number of protojets
unchanged, while merging reduces it by 1.  Therefore there will be
$\order{N}$ split--merge steps before we reach the final list of jets.

There are three kinds of tasks in the split--merge procedure. Firstly
one has to maintain a list of jets ordered in $\pttilde$, both for finding
the one with highest $\pttilde$ and for searching through the remaining
jets (in order of decreasing $\pttilde$) to find an overlapping one.
Maintaining the jets in order is easily accomplished with a balanced
tree (for example a \verb:priority_queue: or \verb:multiset: in C++),
at a cost of $N \ln N$ for the initial construction and $\ln N$ per
update, \ie a total of $N \ln N$, which is small compared to the
remaining steps.

In examining the complexity of finding the hardest overlapping jet one
needs to know the cost of comparing two jets for overlap as well as
the typical number of times this will have to be done. A naive
comparison of two jets takes time $n$. Using a 2d tree structure such
as a quadtree or $k$-d tree (as suggested also by
Volobouev~\cite{Volobouev}), this can be reduced to $\sqrt{n}$. The
number of jets to be compared before an overlap is found will depend
on the event structure --- if one assumes that jet positions are
decorrelated with their $\pttilde$'s, then $\order{N/n}$ comparisons will
have to be made each time around the loop. The total cost of this will
therefore be $N^2/\sqrt{n}$ ($N^2$) with (without) a 2d tree.

Finally each merging/splitting procedure will take $\sqrt{n}$ ($n$)
time with (without) a tree, so the total time spent merging and
splitting will be $\order{N\smash{\sqrt{n}}}$ (or $\order{Nn}$ without
a tree).

The dominant step is the search for overlapping jets, which will have
a total cost of $N^2/\sqrt{n}$ (with a sizable coefficient), or $N^2$
without any 2d tree structures. Since in practice $N^2$ is smaller
than the $N n \ln n$ needed to find the stable cones, here the
introduction of a tree structure gives little overall advantage.

A final comment concerns memory usage: when not using any tree
structures, the list of protojets and their contents requires
$\order{Nn}$ space, which is the same order of magnitude as the
storage needed for identifying the set of stable cones in the first
place. With a tree structure this can be reduced to
$\order{N\smash{\sqrt{n}}}$.

%======================================================================
\section{Proof of IR safety of the SISCone algorithm}
\label{sec:appendix}

In this appendix, we shall explicitly prove that SISCone, algorithms
\ref{alg:fullcone}--\ref{alg:splitmerge}, is infrared safe. This means
that if we run SISCone first with a set of hard particles, then with
the same set of hard particles together with additional soft
particles, then: (a) all jets found in the event without soft
particles will be found also in the event with the soft particles; (b)
any extra jets found in the event with soft particles will themselves
be soft, \ie they will not contain any of the hard particles. If
either of these conditions fails in a finite region of phasespace for
the hard particles, then the cancellation between (soft) real and
virtual diagrams will be broken at some order of perturbation theory,
leading to divergent jet cross sections.

We will first discuss the proof using a simplifying assumption: two
protojets with distinct hard particle content have distinct values for
the split--merge ordering variable, $\pttilde$. 
We shall then discuss subtleties
associated with various ordering variables, and explain why $\pttilde$
is a valid choice.

%----------------------------------------------------------------------
\subsection{General aspects of the proof}
\label{sec:gen-proof}

By soft particles, we understand particles whose momenta 
are negligible compared to the hard ones. Specifically, for any set of
hard particles $\{p_1, \dots, p_n\}$ and any set of soft ones
$\{\bar p_1, \dots, \bar p_m\}$, we consider a limit in which all
soft momenta are scaled to zero, so that they do not affect any
momentum sums,
\begin{equation}\label{eq:momrel}
\lim_{\{\bar p_j\}\to 0} \left(\sum_{i=1}^n p_i + \sum_{j=1}^m \bar p_j\right) =
\sum_{i=1}^n p_i. 
\end{equation}
In what follows, the limit of the momenta of the soft particles being
taken to zero will be implicit.

Let us now compare two different runs of the
cone algorithm: in the first one, referred to as the ``hard event'',
we compute the jets starting with a list of hard particles
$\{p_1,\dots,p_N\}$, and, in the second one, referred to as the
``hard+soft event'', we compute the jets with the same set of hard
particles plus additional soft particles $\{\bar p_1, \dots, \bar
p_{M}\}$. As mentioned above, the IR safety of the SISCone algorithm
amounts to the 
statements (a) that for every jet in the hard event there is a
corresponding jet in the hard+soft event with identical hard particle
content (plus possible extra soft particles) and (b) that there are no hard
jets in the hard+soft event that do not correspond to a jet in
the hard event. To prove this, we shall proceed in two steps: first,
we shall show that the determination of stable cones is IR safe, then
that the split--merge procedure is also IR safe.

The IR safety of the stable-cone determination is a direct consequence
of the fact that:
\begin{itemize}
\item each cone initially built from the hard particles only was
  determined by two particles in algorithm \ref{alg:fastcone}. This
  cone is thus still present when adding soft particles and, because
  of eq.~\eqref{eq:momrel}, is still stable. Hence, all stable cones from
  the hard event are also present after inclusion of soft particles,
  the only difference being that they also contain extra soft
  particles which do not modify their momentum.
\item no new stable cone containing hard particles can appear. Indeed,
  if a new stable cone appeared, $S_{\text{new}}$ with content
  $\{p_{\alpha_1},\dots, p_{\alpha_n}, \bar
  p_{\bar\alpha_1},\dots, \bar p_{\bar\alpha_m}\}$, then the fact
  that its momentum $\sum p_{\alpha_i} + \sum \bar p_{\bar
    \alpha_j}$ corresponds to a stable cone, implies, by
  eq.~(\ref{eq:momrel}), that the cone with just the hard momenta
  $p_{\alpha_i}$ is also stable. However as shown in
  section~\ref{sec:2d} all stable cones in the hard event have already
  been identified, therefore this cone cannot be new.
\end{itemize}
From these two points, one can deduce that after the determination of
the stable cones we end up with two different kinds of stable cones:
firstly, there are those that are the same as in the hard event but
with possible additional soft particles; and secondly there are stable
cones that 
contain only soft particles.  So, the `hard content' of the stable cones
has not been changed upon addition of soft particles and
algorithm~\ref{alg:fastcone} is IR safe. 

The main idea behind the proof of the IR safety of the split--merge
process, algorithm~\ref{alg:splitmerge}, is to show by induction that
the hard content of the protojets evolves in the same way for the
hard and hard+soft event.  Since the hard content is the same at the
beginning of the process, it will remain so all along the split--merge
process which is what we want to prove.

There is however a slight complication here: when running
algorithm~\ref{alg:splitmerge} over one iteration of the loop in the
hard event, we sometimes have to consider more than one iteration
of the loop in the hard+soft event. As we shall shortly see, in that
case, only the last of these iterations modifies the hard content of
the jets and it does so in the same way as in the hard event step.

So, let us now follow the steps of algorithm~\ref{alg:splitmerge} in
parallel for 
the hard and hard+soft event, and show that they are equivalent as
concerns the hard particles. In the following analysis, item numbers coincide with the
corresponding step numbers in algorithm~\ref{alg:splitmerge}.
\begin{itemize}
\item[\ref{alg:splitmerge:threshold}:] If $p_{t,\min}$ is non-zero,
  all purely soft protojets will be removed from the hard+soft event
  and by eq.~\eqref{eq:momrel} the same set of hard protojets will be
  removed in the hard and hard+soft event. Thus the correspondence
  between the hard protojets in the two events will persist
  independently of $p_{t,\min}$. 

\item[\ref{alg:splitmerge:first_jet}:] In general, protojets with
  identical hard content will have nearly identical $\pttilde$ values,
  whereas protojets with different hard-particle content will have
  substantially different $\pttilde$ values.\footnote{As mentioned
    already, this point is more delicate than it might seem at first
    sight. We come back to it in the second part of this appendix.}
  Therefore the addition of soft particles will not destroy the
  $\pttilde$ ordering and the protojet with the largest $\pttilde$ in
  the hard event, $i$ will have the same hard content as the one in
  the hard+soft event (let us call it $i'$).

\item[\ref{alg:splitmerge:second_jet}:] The selection of the
  highest-$\pttilde$ protojet $j$ ($j'$ in the hard+soft case) that
  overlaps with $i$ ($i'$) can differ in the hard and hard+soft
  events, and we need to consider separately the cases where this does
  not, or does happen. 
  The first case, C1, is that $i'$ and $j'$ overlap in their hard
  content --- because of the common $\pttilde$ ordering, $j'$ must then have the
  same hard content as $j$.
  The second case, C2, is that $i'$ and $j'$ only overlap through
  their soft particles, so $j'$ cannot be the `same' jet as $j$ (since
  $j$ by definition overlaps with $i$ through hard particles).
  By following the remaining part of the loop, we shall show that in
  the first case all modifications of the hard content are the same in
  the hard and hard+soft events, while, for the second case, the
  iteration of the loop in the hard+soft event does not modify any
  hard content of the protojets. In this second case, we then proceed
  to the next iteration of the loop in the hard+soft event but stay at
  the same one for the hard event.

\item[{\bf C1}:] The two protojets $i'$ and $j'$ overlap in their hard content
  \begin{itemize}
  \item[\ref{alg:splitmerge:pt_overlap},\ref{alg:splitmerge:start_inner_if}:]
    We need to compute the fraction of $\pttilde$ shared by the two protojets.
    Since the hard contents of $i$ ($j$) and $i'$ ($j'$) are identical,
    the fraction of overlap, given by the hard content only, will be the
    same in the hard and hard+soft events. Hence, the decision to
    split or merge the protojets will be identical.
  \item[\ref{alg:splitmerge:split}:] Since the centres of both protojets are
    the same in the hard and hard+soft events, the decision to
    attribute a hard particle to one protojet or the other will be the same
    in both events. Hence splitting will reorganise hard particles
    in the same way for the hard+soft event as for the hard one.
  \item[\ref{alg:splitmerge:merge}:] In both the hard and the
    hard+soft events, the merging of the two protojets will result in a
    single protojet with the same hard content.
  \end{itemize}

\item[{\bf C2}:] The two protojets $i'$ and $j'$  overlap through soft particles only
\begin{itemize}
\item[\ref{alg:splitmerge:pt_overlap},\ref{alg:splitmerge:start_inner_if}:]
  Since the fraction of $\pttilde$ shared by the protojets will be $0$ in the
  limit eq.~\eqref{eq:momrel}, the two protojets will be split.
\item[\ref{alg:splitmerge:split}:] In the splitting, only shared
  particles, {\em i.e.} soft particles, will be reassigned to the
  first or second protojet. The hard content is therefore left
  untouched, as is the $\pttilde$ ordering of the protojets.
\end{itemize}

\item[\ref{alg:splitmerge:end_inner_if}:] At the end of the
  splitting/merging of the overlapping protojets, we have to consider
  the two possible overlap cases separately: in the first case, the
  hard contents of the protojets are modified in the same way for the
  hard and hard+soft event. This case is thus IR safe.
  In the second case, the iteration of the loop in the hard+soft event
  does not correspond to any iteration of the loop in the hard event.
  However the hard content of the protojets in the hard+soft event is
  not modified and the $\pttilde$ ordering of the jets remains identical;
  at the next iteration of the hard+soft loop, the new $j'$ may once
  again have just soft overlap with $i'$ and the loop will thus
  continue iterating, splitting the soft parts of the jets, but
  leaving the hard content of the jets unchanged. This will continue
  until $j'$ corresponds to the $j$ of the hard event, \ie we
  encounter case 1.\footnote{Note that the second case can only happen
    a finite number of times between two occurrences of the first
    case: as the $\pttilde$ ordering is not modified during the second
    case, each time around the loop the overlap will involve a $j'$
    with a lower $\pttilde$ than in the previous iteration, until one
    reaches the $j'$ that corresponds to $j$.} %
  Therefore even though we may have gone around the loop more times in
  the hard+soft event, we do always reach a stage where the
  split--merge operation in the hard+soft event coincides with that in
  the hard event, and so this part of the procedure is infrared safe.

\item[\ref{alg:splitmerge:test_overlap},\ref{alg:splitmerge:addjet}:] Up to possible intermediate loops
  involving case~2 above, when the protojet $i$ has no overlapping
  protojets in the hard event, the corresponding $i'$ in the hard+soft
  event has no overlaps either.  Final jets will thus be added one by
  one with the same hard content in the hard and hard+soft events.
\end{itemize}
This completes the proof that the SISCone algorithm is IR safe, modulo
subtleties related to the ordering variable, as discussed below.
Regarding the `merge identical protojets' (MIP) procedure:
\begin{itemize}
\item[\ref{alg:splitmerge:disambiguate}:] In
  algorithm~\ref{alg:splitmerge}, we 
  do not automatically merge protojets appearing with the same content
  during the split--merge process. This is IR safe.
  If instead we allow for two identical protojets to be automatically
  merged, then when two protojets have the same hard content
  but differ as a result of their soft content, they are automatically
  merged in the hard event but not in the hard+soft event.
  This in turn leads to IR unsafety of the final jets.
\end{itemize}

A final comment concerns collinear safety and cocircular points. When
defining a candidate cone from a pair of points, if additional points
lie on the edge of the cone, then there is an ambiguity as to whether
they will be included in the cone. From the geometrical point of view,
this special case of cocircular points (on a circle of radius R) can
be treated by considering all permutations of the the cocircular
points being included or excluded from the circle contents. SISCone
contains code to deal with this general issue.  The case of
identically collinear particles,
though a specific example of cocircularity,  also
adds the problem that a circle cannot properly be defined from two
identical points. For explicit collinear safety we thus simply merge any
collinear particles into a single particle,
step~\ref{alg:fastcone:collinear} of algorithm~\ref{alg:fastcone}.
Given the resulting collinear-safe set of  protojets, the split--merge
steps preserve collinear safety, since particles at identical $y-\phi$
coordinates are treated identically.

%----------------------------------------------------------------------
\subsection{Split--merge ordering variable}
\label{sec:split-merge-var}

Suppose we use some generic variable $v$ (which may be $p_t$, $E_t$,
$m_t$, $\pttilde$, etc.) to
decide the order in which we select protojets for the split--merge
process.
A crucial assumption in the proof of IR safety is that two jets with
different hard content will also have substantially different values
for $v$, \ie the ordering of the $v$'s will not be changed by soft
modifications. If this is not the case then the choice of the hard
protojets that enter a given split--merge loop iteration can be
modified by soft momenta, with a high likelihood that the final jets
will also be modified.

At first sight one might think that whatever variable is used, it will
have different values for distinct hard protojets. However, momentum
conservation and coincident masses of identical particles can
introduce relations between the kinematic characteristics of distinct
protojets. Some care is therefore needed so as to ensure that these
relations do not lead to degeneracies in the ordering, with consequent
ambiguities and infrared unsafety for the final jets. In particular:

\begin{itemize}
\item Two protojets can have equal and opposite transverse momenta if
  between them they contain all particles in the event (and the event
  has no missing energy or `ignored' particles such as isolated
  leptons). It is probably fair to assume that no two protojets will
  have identical longitudinal components, since in $pp$ collisions the
  hard partonic reaction does not occur in the $pp$ centre of mass
  frame.

\item Two protojets will have identical masses if they each stem
  exclusively from the same kind of massive particle. The two massive
  particles may be undecayed (\eg fully reconstructed $b$-hadrons) or
  decayed (top, $W$, $Z$, $H$, or some non-standard new particle), or
  even one decayed and the other not (some hypothetical particle with
  a long lifetime).\footnote{Strictly speaking, for all scenarios of
    decayed heavy particles, the finite width $\Gamma$ of the particle
    ensures that the two jets actually have slightly different masses,
    breaking any degeneracies. In practice however, $\Gamma_{W,Z,t}
    \sim 1\GeV$ and (for a light Higgs) $\Gamma_H \ll \Lambda_{QCD}$,
    whereas for the width to save us from the dangers of degeneracies
    we would need $\Gamma \gg \Lambda_{QCD}$.} %
  In the second case we can assume that two identical decayed particles
  have different decay planes, because there is a vanishing phase space for
  them to have identical decay planes.
\end{itemize}
Note that in a simple two-parton event almost any choice of variable
will lead to a degeneracy (no sensible invariant will distinguish the
two particles), however this specific case is not problematic because
for $R<\pi/2$ neither of the two partons can be in a protojet that
overlaps with anything else. From the point of view of IR safety, it
is only for `fat' (non-collimated) hard protojets that we need worry about
the problem of degeneracies in the split--merge ordering, because only
then will there be overlaps whose resolution is ambiguous in the
presence of degeneracies.

Let us now consider what occurs with various possible choices for the
split--merge variable.
\begin{itemize}
\item[$p_t$:] This choice, adopted in certain
  codes~\cite{CDFcode,NLOJet}, can be seen to have a problem for
  events with momentum conservation in the hadronic part, because if
  two non-overlapping protojets contain, between them, all the hard
  particles then they will have identical $p_t$'s. If they each
  overlap with a common third protojet, the resulting split--merge
  sequence will be ambiguous. Table~\ref{table:sm-pt-problem} provides
  an example of such an event.
  The simplest occurrences of this problem ($4h+1s$) apply only to
  $R>\pi/4$ (four particles must form at least 3 fat protojets). The
  problem arises also for smaller $R$ values, but only at higher
  multiplicities.

\item[$m_t$:] A workaround for the event of
  table~\ref{table:sm-pt-problem} is to use the transverse mass, $m_t
  = \sqrt{p_t^2 + m^2}$. In pure QCD, with all particles stable, this
  is a good variable, because even if two fat protojets have identical
  $p_t$'s through momentum conservation, the fact that they are `fat'
  implies that they will be massive (over and above intrinsic particle
  masses), and the phase space for them to have identical masses
  vanishes, thus killing any IR divergences.
  However, for events with two identical decaying particles, two fat
  protojets resulting from the particle decays can have identical
  $p_t$'s (by momentum conservation) and identical masses (because the
  decaying particles were identical). This could happen for example in
  the fully hadronic decay channel for $t\bar t$ events. Thus, this
  choice is not advisable in a general purpose algorithm.

\item[$E_t$:] The variable used in the original run~II proposal was
  $E_t$~\cite{RunII-jet-physics}. It has the drawback that it is not
  longitudinally boost invariant: at central rapidity it is equal to
  $m_t$, while at high rapidities it tends to $p_t$. Because the phase
  space for two protojets to have identical rapidities vanishes
  (recall that we do not fix the partonic centre-of-mass), two
  protojets with identical $p_t$'s and masses will have different
  $E_t$'s, because the degree of `interpolation' between between $p_t$
  and $m_t$ will be different. This resolves the degeneracy and should
  cure the resulting IR safety issue, albeit at the expense of
  introducing boost-dependence.

\item[$\pttilde$:] The scalar sum of transverse momenta of the
  protojet constituents, $\pttilde$, has the property that it is equal
  to $m_t$ if all particles in the protojet have identical rapidities,
  while it is equal to $p_t$ (\ie the vector sum) if all particles
  have identical azimuths. For a decayed massive particle, it
  essentially interpolates between $p_t$ and $m_t$ according to the
  orientation of the decay plane.
  The phase space for all particles to have identical azimuths
  vanishes, as does the phase space for the decay products of two
  heavy particles to have identically oriented decay planes.
  Therefore this choice resolves any degeneracies, as is needed for
  infrared safety.  Another advantage of $\pttilde$ is that adding a
  particle to a protojet always increases its $\pttilde$ (this is not
  the case for $p_t$ or $E_t$), ensuring that the degree of overlap
  between a pair of jets is always bounded by $1$.
  Since it is also boost invariant, it is the choice that we
  recommend and that we adopt as our default.\footnote{%
    One might worry about the naturalness of a variable that depends
    on the decay plane of heavy particles --- however, any
    unnaturalness is present anyway in the split--merge procedure
    since if two particles decay purely in the transverse plane then
    there is a likelihood of having overlapping protojets, whereas if
    they decay in longitudinally oriented decay planes they will not
    overlap.} %
\end{itemize}
%=============================
% illustrative event 
\begin{table}
  \begin{center}
    \begin{tabular}{c|rrr}
      \multicolumn{4}{c}{event 1}\\\hline
      n &  $p_x$ & $p_y$ & $p_z$\\\hline
      0 &   86.01 &   66  & 0   \\ 
      1 &   64   &  -66  & 0    \\
      2 &  -77   &  -70  & 0    \\
      3 &  -73   &   70  & 0    \\
      4 &  -0.01  &   0  & 2   \\
    \end{tabular}\qquad\qquad
    \begin{tabular}{c|rrr}
      \multicolumn{4}{c}{event 2}\\\hline
      n &  $p_x$ & $p_y$ & $p_z$\\\hline
      0 &   85.99 &   66  & 0   \\ 
      1 &   64   &  -66  & 0    \\
      2 &  -77   &  -70  & 0    \\
      3 &  -73   &   70  & 0    \\
      4 &   0.01  &   0  & 2   \\
    \end{tabular}
    \caption{Illustration of two events that conserve transverse momentum and
      differ only through a soft particle, but lead to different hard
      jets  with a split--merge procedures that uses $p_t$ as the
      ordering variable and for measuring overlap. All the particles
      are to be taken massless.
      For $R=0.9$ and $f=0.7$ each event has stable cones consisting
      of $\{01\}$, $\{23\}$ and $\{12\}$, as well as all single
      particles. The slight difference in momenta between the two
      events, to balance the soft particle, causes the $\{01\}$
      ($\{23\}$) protojet to have the largest $p_t$ in the first
      (second) event, it splits with $\{12\}$ (merges with $\{12\}$),
      leading after further split--merge steps to two hard jets,
      $\{01\}$ and $\{23\}$ (one hard `monster' jet, $\{0123\}$).
      \label{table:sm-pt-problem}
    }
  \end{center}
\end{table}
Note that the above considerations hold for any split--merge procedure
that relies on ordering the jets according to a single-jet variable.
One might also consider ordering according to variables determined
from pairs of protojets: e.g.  first split-merge the pair of protojets
with the largest (or alternatively smallest) overlap, recalculate all
overlaps, and then repeat until there are no further overlaps. However
this specific example would also be dangerous, since the particles that
are common to protojets $a$ and $b$ (say) could also be the particles
that are common between $a$ and $c$, once again leading to an
ambiguous split--merge sequence. One protojet-pair ordering variable
that might be free of this problem is the $y-\phi$ distance between
the protojets, however we have not investigated it in detail.

A final comment concerns the impact of the split--merge procedure on
non-global~\cite{non-global} resummations for
jets~\cite{non-global-jets}, in which one is interested in determining
which of a set of ordered soft particles are in a given hard jet.  A
soft and collinear splitting inside the jet can modify the $\pttilde$
(or $E_t$ or $m_t$) of the jet by an amount of the same order of
magnitude as a soft, large-angle emission near the edge of the jet. In
events with two back-to-back narrow jets, for which there is a near
degeneracy between the $\pttilde$'s of the two hard jets, this can
affect which of the two hard protojets split--merges first with an
overlapping soft protojet, leading to ambiguities in the assignment of
the soft particles to the two hard jets. This interaction between
collinear and soft modes is somewhat reminiscent of that
in~\cite{FKS}, though the origin and structure are kinematical
in our case.
Considering only branchings with transverse momenta above
$\epsilon p_{t,\mathrm{hard}}$, for $R>\pi/4$ this is likely to be
relevant in events with two equally soft particles ($\as^2 \ln
\epsilon$) and $n$ soft-collinear splittings ($\as^n \ln^{2n}
\epsilon$) giving an overall contribution $\as^{n+2} \ln^{2n+1}
\epsilon$. This competes with the normal soft-ordered non-global
logarithms, starting from order $\as^{3} \ln^{3} \epsilon$.  For $R\le
\pi/4$, the problem will only arise with a greater number of equally
soft large-angle particles, and so will be further suppressed by
powers of $\as$.

\end{appendix}

\end{document}